\documentclass[preprint2]{aastex}
\usepackage{graphics}
\usepackage{epsfig}

\def\chandra{{\sl Chandra}}
\newcommand{\xmm}{{\em XMM-Newton}}

\def\hst{{\sl HST}}

\newcommand{\as}{$^{\prime\prime}$}

\def\xs{Arches}
\def\xq{Quintuplet}

\shortauthors{}
\shorttitle{Chandra Observation of \xs}

\begin{document}
\slugcomment{Draft version: \today}

\title{\bf The Interplay between Star Formation and the Nuclear Environment of our Galaxy: Deep X-ray Observations of the Galactic Center Arches and Quintuplet Clusters}

\author{Q. Daniel Wang\altaffilmark{1}, Hui Dong\altaffilmark{1}, \& Cornelia Lang\altaffilmark{2}}
\altaffiltext{1}{Department of Astronomy, B619E-LGRT,
       University of Massachusetts, Amherst, MA~01003; wqd@astro.umass.edu, hdong@astro.umass.edu}
\altaffiltext{2}{Department of Physics and Astronomy, University of Iowa, Iowa City, IA 52245; cornelia-lang@uiowa.edu}

\begin{abstract}
The Galactic center (GC) provides a unique laboratory for a
detailed examination of the interplay between massive star
formation and the nuclear environment of our Galaxy. Here, we 
present an 100 ks {\sl Chandra} ACIS observation of the Arches and Quintuplet star clusters. We
also report on a complementary mapping of the dense molecular 
gas near the Arches cluster made with the Owens Valley Millimeter Array. 
We present a catalog of 244 point-like X-ray sources detected in the observation. 
Their number-flux relation indicates an
over-population of relatively bright X-ray sources, which 
are apparently associated with the clusters. The sources in the core of the 
Arches and Quintuplet clusters are most likely extreme colliding wind
massive star binaries. The diffuse X-ray emission from the core
of the Arches cluster has a spectrum showing a 6.7-keV emission line and a
surface intensity profile declining steeply with radius, indicating an
origin in a cluster wind. In the outer regions near the Arches cluster, the
overall diffuse X-ray enhancement demonstrates a bow shock morphology and is prominent
in the Fe K$\alpha$ 6.4-keV line emission with an equivalent width
of $\sim 1.4$ keV. Much of this enhancement may result from an ongoing collision
between the cluster and the adjacent molecular cloud, which have a relative velocity
$\gtrsim 120 {\rm~km^{-1}}$.  The older and less
compact Quintuplet cluster contains much weaker X-ray sources and
diffuse emission, probably originating from low-mass stellar objects as well as a cluster 
wind.  However, the overall population of these objects, constrained by the observed 
{\sl total} diffuse 
X-ray luminosities, is substantially smaller than expected for both clusters, if they have 
normal Miller \& Scalo initial mass functions. This deficiency of low-mass objects 
may be a manifestation of the unique star formation environment of the Galactic center,
where high-velocity cloud-cloud and cloud-cluster collisions are frequent.
\end{abstract}


\section{Introduction}\label{s:intr}

Nuclear regions of galaxies are the breeding ground of high energy phenomena
and processes, which are manifested observationally by active galactic nuclei (AGNs) and
star-bursts. Such activities are believed to be important
in both regulating galaxy evolution and generating
thermal and chemical feedback into the intergalactic medium.  The best
site for a detailed study of the activities and their complex
interaction with the physically extreme environment in the nuclear regions of
galaxies is our own nucleus, only $\sim 8$ kpc away. We can observe the Galactic center (GC)
with a spatial resolution and sensitivity that are
factors $\gtrsim 300$ and $ \gtrsim 10^5$ better than those
available for even nearby nuclear starburst galaxies (e.g., M82 and NGC 253) or
AGNs (e.g., M81).

While the super-massive black hole at the dynamic center of the
Galaxy is only weakly active at present
\citep{bag01}, much of the current high-energy activity in the GC is due to the
presence of the three young massive stellar clusters in the central 50 pc: Arches (with age
equal to $2-3 \times 10^6$ yrs; core or half-mass radius 0.4 pc),
Quintuplet ($3-6 \times 10^6$ yrs; 1.0 pc), and the Central cluster ($3-7 \times
10^6$ yrs; 0.5 pc)
(Figer et al. 1999, 2002, 2004; Stolte et al. 2002; Genzel et al. 2003).
These clusters are
responsible for about half of the Lyman continuum flux emitted in
the central several $10^2$ pc of the GC. Massive stars are also
expected to release large amounts of mechanical energy into the GC
region in form of fast stellar winds and supernovae, although the
actual rate is highly uncertain. This mechanical energy input
shapes the surrounding ISM. The present work focuses on
the \xs\ and \xq\ clusters. The GC cluster is less massive and
older. Its location in the circum-nuclear region makes its X-ray properties 
difficult to characterize and will not be dealt with here (however see \citealt{nay05}).

Both the \xs\ and \xq\ clusters are known X-ray emitters. Discovered
serendipitously at a large off-axis angle ($\sim 7^\prime$) in an
initial 50 ks {\sl Chandra} ACIS-I observation
\citep{yus02}, the X-ray emission arising from the \xs\ cluster was resolved into discrete and diffuse components. In a later 12 ks ACIS-I observation
during a large-scale GC survey \citep{wan02a}, the X-ray core of the cluster was further resolved into two
separate components \citep{wan03,law04}. The apparent diffuse X-ray component was speculated to arise
from the so-called cluster wind \citep{rag01,yus02}. Because of the high number density of massive stars, their stellar
winds collide with each other and can be largely thermalized to form a plasma with an
initial temperature of a few times $10^7$ K. The expanding of this
plasma may then be considered as a wind from the entire cluster.
However, the quality of the previous observations is not adequate
for a quantitative test of this scenario. In particular, the
diffuse X-ray spectrum shows a distinct 6.4-keV emission line from
neutral or weakly ionized irons. The origin of this line is
unknown. The X-ray emission from the \xq\ cluster is substantially
weaker. The detection of a few discrete sources and possible
diffuse emission in the region has been reported; but detailed
spectral and timing information is not yet available
\citep{wan03,law04}

To further the study of these two clusters and their relationship to the
environment, we have obtained
an 100 ks \chandra\ ACIS-I observation.
We have further carried out a complementary imaging study of the molecular
gas in the immediate surroundings of the Arches cluster using the six-element
Owens Valley Millimeter Array.  With these observations and other
multi-wavelength data, we present an in-depth study of
various point-like and diffuse X-ray sources in and around the
clusters. 

In this paper, we assume that the distance to the GC is 8
kpc (hence $1^\prime =$ 2.5 pc) and quote statistical errors from
our X-ray data analysis at the 90\% confidence level, unless being
pointed out otherwise. The solar abundance is in reference to
\citet{and89}; thus the number of iron relative to hydrogen is $4.68 \times
10^{-5}$,  which is considerably greater than $2.69 \times
10^{-5}$ in the so-called ISM abundance \citep{wil02}, for example.

\section{Observations and Data Reduction}

\subsection{\chandra\ Observations}\label{ss:chandra}

Our deep \chandra\ ACIS-I observation (Obs. ID:  4500) was carried out on June 9, 2004.
The \xs\ cluster was placed about 1$^\prime$ away from the aim-point to minimize the
effect of the CCD gaps on mapping the extended X-ray emission around
the cluster. This slight offset from the axis had a
negligible effect on the spatial resolution for the \xs\ cluster.
We used the ``very faint'' mode for a better discrimination and removal of
charged particle induced events.  We have reprocessed the data, using the
CIAO software (version 3.2.1) and calibration database
(version 3.0.0). This reprocessing includes both  charge
transfer inefficiency (CTI) and gain corrections as well as the
removal of time intervals contaminated by background flares.
The total reprocessed good time (live-time) is 98.6 ks.
We create ACIS-I event images and corresponding exposure
maps in the 1-2.5, 2.5-4, 4-6, and 6-9 keV bands.

We detect X-ray sources, following the same procedure as detailed
in \citet{wan04}. The detection, optimized for
point-like sources, uses a combination of algorithms: wavelet,
sliding-box, and maximum likelihood centroid fitting. The source
detections are carried out in the 1-4
keV, 4-9 keV, and 1-9 keV bands. The detected sources in the
three bands are merged together. Multiple detections with
overlapping 2$\sigma$ centroid error circles are considered to be
the same source, and the centroid with the smallest error is
adopted. The accepted source candidates generally have individual
local false detection probabilities $P\le 10^{-7}$. But in the vicinity
($2^\prime \times 2^\prime$ field) of the clusters, we also
include sources detected with reduced significances
 $10^{-7} < P\le 10^{-5}$. Over the entire search, the expected false detection probability is
$\sim 1$.

We check the astrometry of the X-ray observation, based on the
multi-wavelength comparisons. A SIMBAD search gives some potential
counterparts within the 3$\sigma$ error radius around each
detected X-ray source. A few sources in the \xs\ cluster have
radio counterparts, which have accurate positions ($\sim 0\farcs1$; 
\citealt{lan05}. A comparison of the two bright X-ray/radio pairs 
(92/AR1 and 93/AR4; Table~\ref{t:sou_cl}) shows that their positions
are all within $0\farcs3$, consistent with their statistical and systematic 
uncertainties (Table~\ref{t:sou}). Finding exact matches of X-ray sources
with near-IR objects is generally more difficult, because they are numerous
and typically have relatively large absolute position
uncertainties of $\sim 1^{\prime\prime}-2^{\prime\prime}$. We approximately
correct for the  relatively shifts of the near-IR
positions to the X-ray positions, using the original \hst\ NICMOS
observations of the \xs\ and Quintuplet clusters as well as the
coordinates of individual NIR objects (2, 6, 7, and 9 for the \xs\ and
211, 242, 231, and 257 for the \xq) listed in
\citep{fig99a,fig02}. The {\sl
HST}-to-\chandra\ RA. and Dec. shifts are -0\farcs52 and 0\farcs29
for the Arches cluster and -0\farcs43 and -0\farcs32 for the Quintuplet
cluster, respectively. These shifts are then applied to the NICMOS
images to facilitate the comparison with the X-ray data. The uncertainties
in these astrometry corrections should be around $0\farcs3$, dominated by
the errors in the X-ray source centroids.

To construct ``diffuse'' X-ray maps, we excise the detected sources from
the ACIS-I data. For each source with a count rate (CR) $\leq 0.01
{\rm~counts~s^{-1}}$, we exclude a circular region with a radius
of 1.5 times the 90\% PSF energy enclosed radius (EER).
For sources with CR $>$ 0.01 cts/s, this radius is multiplied by
an additional factor of 1+log(CR/0.01) to further minimize the contamination
from the PSF wing.

For the background subtraction in our imaging analysis,
we use the blank-sky data with a total
exposure of 550 ks. The data are re-projected to mimic individual observations.
The background subtraction is mostly to remove the contribution from
events induced by charged particles.
Of course, the blank-sky data also contains cosmic X-ray
radiation. Its intensity varies from one part of the sky to another, mostly
at energies below $\sim 1$ keV. At higher energies, the radiation is 
negligible, compared to the fluxes due to charged particle-induced 
events and to the emission from the GC region. The combination of
the background subtraction and the exposure correction then gives
the flat-fielded intensity images in individual bands.

In addition to the broad-band images, we also construct narrow-band images
of the prominent 6.4-keV and 6.7-keV emission
lines in the energy ranges of 6.25-6.55 keV and 6.55-7 keV. Because the counting statistics at energies
greater than 7 keV is too poor, we estimate the
continuum contribution to these narrow bands, based on the intensities measured
in the 4-5 keV and 5-6.2 keV bands. We assume an intrinsic
power law spectral shape of the blank-sky background-subtracted
diffuse emission in the 4-7 keV range and account for both
the effective area and energy response of the instrument,
using the convolved model outputs from the X-ray spectral analysis
software {\small XSPEC}. The continuum-subtracted line intensity image, divided by the
specific continuum intensity, gives the equivalent width (EW) map of the line.
The calculation at each image pixel is carried out adaptively, using a Gaussian kernel
with its size adjusted to achieve a signal-to-noise ratio greater than five at each step.

We further include all available archival ACIS-I observations that were
taken before our deep observation and covered
the clusters \citep{yus02,wan02a} in analyzes that do not require
the maximum spatial resolution offered by our on-axis observation.
The combined data have an effective exposure of 157 ks at \xs\ and
160 ks at \xq. Source detections are also carried out for these
individual observations and are used to examine the potential
long-term variability of the sources in the close vicinity of the
clusters.
For detailed spatial and spectral analyzes of the \xs\ cluster, we
use only our on-axis deep \chandra\ observation. But for the \xq\ cluster,
all the observations are off-axis and and are thus used.
Because the effective area and spectral response depend on both
time and position, we extract spectra from individual observations
separately and then combine them together using the {\small
FTOOLS} routine ``addspec'', which produces weighted effective
area and spectral response files. All spectral extractions use the
``Gaussian error'', not the ``Poisson error'', which is the
default of the {\small CIAO} {\sl psextract} routine and actually
uses the Gehrels's approximation \citep{geh86}. This latter method could cause problems in the error
propagation through the spectral co-adding. For bright X-ray
point-like sources in the \xs\ cluster, we typically group such spectra to
achieve a minimum 20 counts per bin. For diffuse X-ray emission
spectra, the noise contribution from the subtracted background
becomes important. We group the spectra to have the {\sl
background-subtracted} signal-to-noise (S/N) in each bin greater
than 3 for the \xs\ cluster and 2 for the \xq\ cluster with
fainter diffuse X-ray emission.

\subsection{CS (J=2-1) Molecular Line Observations}\label{ss:cs}

The region around the GC Radio Arc (where the Quintuplet and Arches clusters
are located) harbors a number of large, dense molecular clouds. The so-called 
``$-30 {\rm~km~s^{-1}}$ cloud'' is believed to be ionized by the Arches cluster \citep{lan01a,lan02}. High resolution ($\sim$ 10\as) observations of part of this
cloud complex which immediately surrounds the Arches cluster 
were made in the 3.4 mm continuum and CS (J=2-1)
line using the six-element millimeter array at the Owens Valley Radio Observatory (OVRO) in March,
April, May and June 2002. Two telescope configurations (equatorial and low) were used, with baselines
ranging from 15 to 100 m. Six fields with a primary beam of 60\as\ were observed in a mosaic pattern,
with a spacing of 30\as. The resulting mosaic covers an area of approximately 4$^\prime \times 3^\prime$ centered on the
position $RA=17^h 46^m 51^s, DEC=-28^\circ 49^{\prime} 00^{\prime\prime}$ (J2000). The total integration time on each of the six fields was approximately 4 hours.

NRAO 530, 3C 273, and Neptune were used for gain, passband, and absolute flux calibration,
respectively. The data were calibrated using the {\small MMA} package (Scoville et al. 1993), and
the mosaicking was carried out with the maximum entropy method of de-convolution implemented
in the {\small MIRIAD} routine MOSMEM (Cornwell \& Braun 1988; Sault, Staveley-Smith, \& Brouw 1996). The
CS (J=2-1) line data were taken at a rest frequency of 97.981 GHz, with 64 channels of 0.5 MHz width,
corresponding to a velocity resolution of 1.53 ${\rm~km~s^{-1}}$ and a total velocity coverage of 96 ${\rm~km~s^{-1}}$. The
line was centered on $v_{LSR}=-20$ ${\rm~km~s^{-1}}$. Simultaneous 3.4 mm continuum observations were made with a
bandwidth of 1 GHz.

The largest spatial scale to which the OVRO interferometer is sensitive is 20\as, corresponding
to the shortest baseline of 15 m at 3.4 mm. Therefore, more extended structures are not detected.
In order to recover the missing flux density, the total power measurements from single-dish
observations of this region have been added. Single-dish observations of the CS (2-1) line in
the -30 ${\rm~km~s^{-1}}$ molecular cloud were carried out with the {\it IRAM} 30 m telescope by \citep{ser87}.
Spectra in the vicinity of the Arched Filaments complex were obtained at regular grid spacings of 18\as\
and imaged with a single-dish beam size of 25\as. These observations were centered at $v_{LSR} = 0$ ${\rm~km~s^{-1}}$,
using a 512-channel filter bank with 1 MHz resolution, which corresponds to a velocity resolution of 3.06 ${\rm~km~s^{-1}}$.

Since there is reasonable overlap between the shortest spacings of the OVRO interferometer
(4 kilo-lambda) and the diameter of the 30 m antenna (8 kilo-lambda), the linear technique of
``feathering'' single-dish and interferometer data is appropriate. This method requires that the
single-dish data be a good representation of the object at low spatial frequencies, and that the
interferometer mosaic be a good representation at mid-to-high spatial frequencies. The feathering
technique can be carried out using the MIRIAD task IMMERGE. We input de-convolved and restored
single-dish and interferometric images with the same velocity resolution and spatial grid. IMMERGE
first transforms the images into the Fourier plane, where the data are combined. In the case of the 30 m single dish data and OVRO millimeter array data, the flux densities in the overlap region (4-8 kilolambda) agree at the 10\% level. The single-dish data are given unit weight, and the low spatial frequencies of the interferometer data are adjusted
in the Fourier plane with a taper such that a combination of the single-dish and interferometer
data results in an image with a Gaussian beam equal in diameter to the beam of the input interferometer
mosaic image ($\sim$9\arcsec~$\times$9\arcsec). 

\section{Analysis and Results}\label{s:results}

\begin{figure*}[!thb]
  \centerline{
      \epsfig{figure=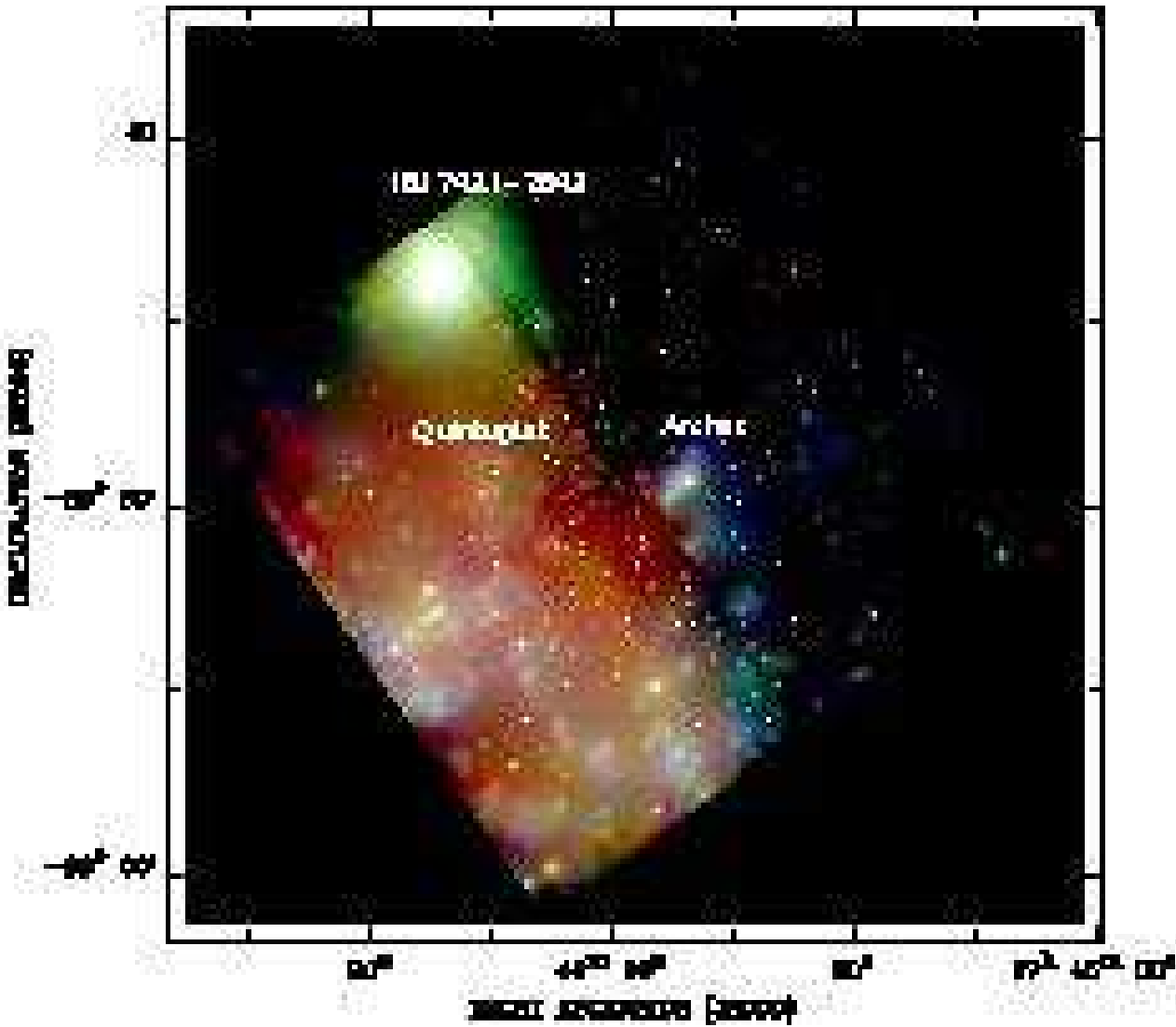,width=1.1\textwidth,angle=0}
    }
  \caption{\small Tri-color presentation of the 100 ks {\sl Chandra} ACIS-I observation. Artifacts due to
the gaps between the four CCDs and to their outer edge are still visible and are partly caused by
sharp changes in the counting statistics.}
\label{f:ob_rgb}
\end{figure*}

The entire field of our deep ACIS-I observation is
shown in Fig.~\ref{f:ob_rgb}, while Fig.~\ref{f:im_multi_bw} gives 
close-ups of the \xs\ and \xq\ clusters.
In this section, we first examine the overall X-ray source population in the
field of the observation and then present a detailed characterization of
 discrete sources and diffuse emission in the \xs\ and \xq\ regions,
separately.

\begin{deluxetable}{lrrrrrrrr}
  \tabletypesize{\footnotesize}
  \tablecaption{{\sl Chandra} Source List \label{acis_source_list}}
  \tablewidth{0pt}
  \tablehead{
  \colhead{Source} &
  \colhead{CXGCS Name} &
  \colhead{$\delta_x$ ($''$)} &
  \colhead{CR $({\rm~cts~ks}^{-1})$} &
  \colhead{HR} &
  \colhead{HR2} &
  \colhead{Flag} \\
  \noalign{\smallskip}
  \colhead{(1)} &
  \colhead{(2)} &
  \colhead{(3)} &
  \colhead{(4)} &
  \colhead{(5)} &
  \colhead{(6)} &
  \colhead{(7)}
  }
  \startdata
  85 &  J174549.73-284926.1 &  0.3 &$     0.43  \pm   0.08$& --& -- & B, H \\
  90 &  J174550.26-284911.9 &  0.2 &$     7.48  \pm   0.29$& $-0.32\pm0.05$ & $-0.30\pm0.06$ &B, S, H \\
  92 &  J174550.41-284922.4 &  0.2 &$    11.03  \pm   0.36$& $-0.34\pm0.04$ & $-0.40\pm0.05$ &B, S, H \\
  93 &  J174550.47-284919.7 &  0.2 &$     6.90  \pm   0.29$& $-0.39\pm0.05$ & $-0.50\pm0.06$ &B, S, H \\
 213 &  J174614.44-284908.6 &  0.6 &$     1.49  \pm   0.14$& $ 0.27\pm0.12$ & $-0.65\pm0.09$ &B, H, S \\
 214 &  J174614.51-284937.2 &  0.6 &$     0.70  \pm   0.11$& $-0.81\pm0.12$ & -- & B, S \\
 215 &  J174614.67-284940.3 &  0.7 &$     0.34  \pm   0.09$& --& -- & B, S \\
 216 &  J174615.14-284932.9 &  0.7 &$     0.50  \pm   0.10$& --& -- & B, S \\
 217 &  J174615.85-284945.5 &  0.8 &$     0.57  \pm   0.10$& --& -- & B, H \\
 219 &  J174616.29-284940.8 &  0.8 &$     0.34  \pm   0.09$& --& -- & B, H \\
\\
\multispan{7}{\hfill Sources detected with $10^{-7} < P < 10^{-5}$ and in the vicinity of the \xs\ and \xq\ clusters \hfill }\\
\\
 240 &  J174549.35-284919.0 &  0.4 &$     0.16  \pm   0.05$& --& --& B \\
 243 &  J174614.43-284900.0 &  0.7 &$     0.25  \pm   0.08$& --& --& S \\
 244 &  J174616.66-284909.2 &  0.7 &$     0.26  \pm   0.08$& --& --& B \\
\enddata
\tablecomments{The printed version of the table includes only the sources
within the {\sl HST} NICMOS fields of the \xs\ and \xq\ clusters
(Fig.~\ref{f:im_multi_bw}s c and d); the 
full source list is published only electronically. The definition of the bands:
1--2.5 (S1), 2.5--4 (S2), 4--6 (H1), and 6--9~keV  (H2). 
In addition, S=S1+S2, H=H1+H2, and B=S+H.
 Column (1): Generic source number. (2):
{\sl Chandra} X-ray Observatory (registered) source name, following the
{\sl Chandra} naming convention and the IAU Recommendation for Nomenclature
(e.g., http://cdsweb.u-strasbg.fr/iau-spec.html). (3): Position
uncertainty, including an 1$\sigma$ statistical error calculated from
the maximum likelihood centroiding and an approximate off-axis angle ($r$)
dependent systematic error $0\farcs2+1\farcs4(r/8^\prime)^2$
(an approximation to Fig.~4 of \citet{fei02}), which are added in
quadrature. 
(4): On-axis source broad-band count rate --- the sum of the
exposure-corrected count rates in the four
bands. (5-6): The hardness ratios defined as
${\rm HR}=({\rm H-S2})/({\rm H+S2})$, and ${\rm HR2}=({\rm H2-H1})/{\rm H}$, 
listed only for values with uncertainties less than 0.2.
(7): The labels ``B'', ``S'', and/or ``H'' mark the bands in
which a source is detected; the band which generates the most
accurate X-ray centroid position, as adopted in Column (2), is listed first.
}
\label{t:sou}
\end{deluxetable}

\subsection{Discrete X-ray Sources}\label{ss:sou_pop}

Table~\ref{t:sou} summarizes the key parameters of our detected X-ray 
sources in the deep ACIS-I observation.
The note to this table explains various parameters listed.
The hardness ratios, in particular, provide simple source spectral
characteristics, which may be compared with model predictions (e.g., Fig.~\ref{f:hr}).
A few sources show exceptionally large HR2 values, which could be
reproduced with the assumed models only with abnormal parameters
(e.g., a power law with a photon index $\Gamma < 0$ or a plasma with an iron abundance
$> 2\times$solar). Assuming a plasma with a higher temperature ($\gtrsim 6$ keV)
would not help, which reduces the He-like Fe line emission and hence
the HR2 value.

\begin{figure*}[!thb]
  \centerline{
      \epsfig{figure=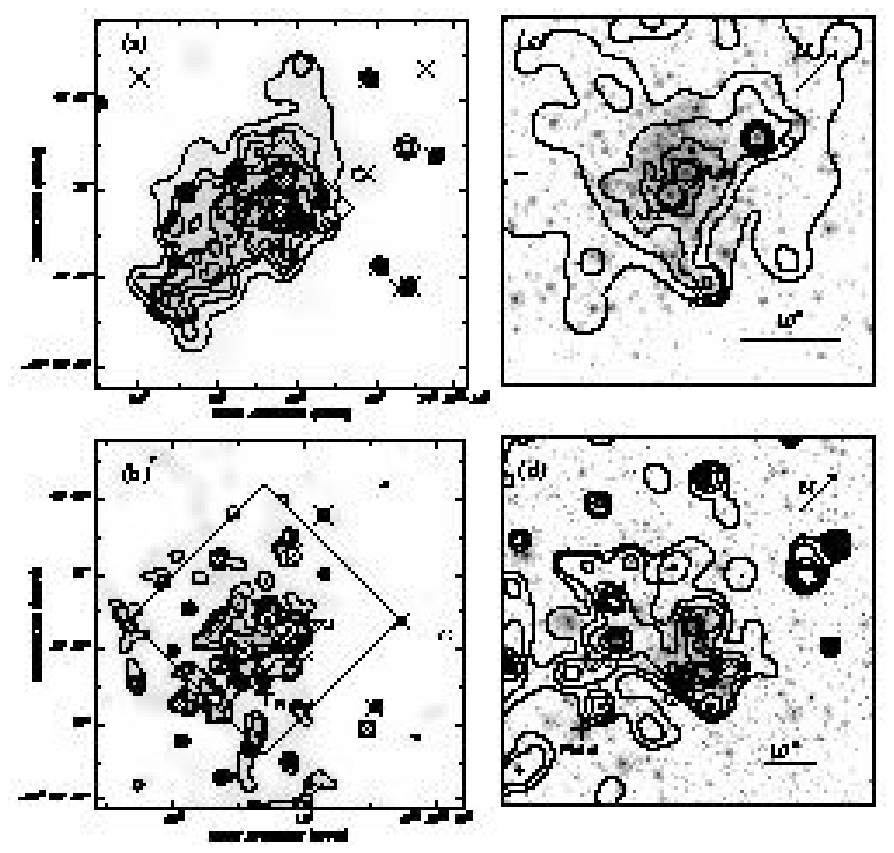,width=1.05\textwidth,angle=0}
    }
  \caption{\small ACIS-I 1-9 keV band images of the \xs\ (a) and
\xq\ (b) clusters. These X-ray images are smoothed with the
{\small CIAO} routine {\sl csmooth} to
achieve a background-subtracted signal-to-noise ratio of $\sim 3$
\citep{ebe06}. The intensity contour levels
are at 20, 23, 27, 33, 43, 57, 80, 180,
482, and 1351 (above a local background of 13.4)
for (a), at 17, 29, 33, 42, 54, and 72 (above 17) for
(b); all in units of $10^{-3} {\rm~counts~s^{-1}~arcmin^{-2}}$.
The two large squares in (a) and (b) outline the fields covered by
the {\sl HST} NICMOS F205W images of the \xs\ (c) and \xq\ (d),
respectively \citep{fig99a,fig02}. The contours are the same as in (a) and (b),
except for excluding the first four levels in (b).
The detected sources (Table~\ref{t:sou})
are marked with {\sl crosses} in (a) and (b). Several bright X-ray
sources named previously (Table~\ref{t:sou_cl}) are labeled.
}
\label{f:im_multi_bw}
\end{figure*}

We check timing variability for each of the sources,  based on
a Kolmogorov-Smirnov test with the light curve of the
source-removed background as
a reference. The test is performed in all three source detection energy bands.
Only Source \#20 (J174532.27-285052.3) in the table shows marginal evidence ($3\sigma$)
for variability.

The brightest source in the field is LMXB 1E 1743.1-2843 with
an X-ray luminosity of  $L_X\sim 2 \times 10^{36} {\rm~ergs~s^{-1}}$ in the 2-10~keV
band (assuming at the distance of the GC; \citealt{por03}). The
second brightest source is apparently a renewed X-ray burst of XMMU
J174554.4-285456 \citep{por05}, detected during an \xmm\
observation
performed on October 3, 2002. This source did {\sl not} appear in a
\chandra\ observation (OBSID \#3549, June, 19, 2003) in-between the \xmm\
observation and our detection here (June 9, 2004).
The source was also present in the two
subsequent \chandra\ observations (OBSID \#4683, July 5, 2004 and \#4684,
July 6, 2004), but not in later ones (e.g., OBSID \# 5360, August 28, 2004).

\begin{figure*}
\centerline{
\psfig{figure=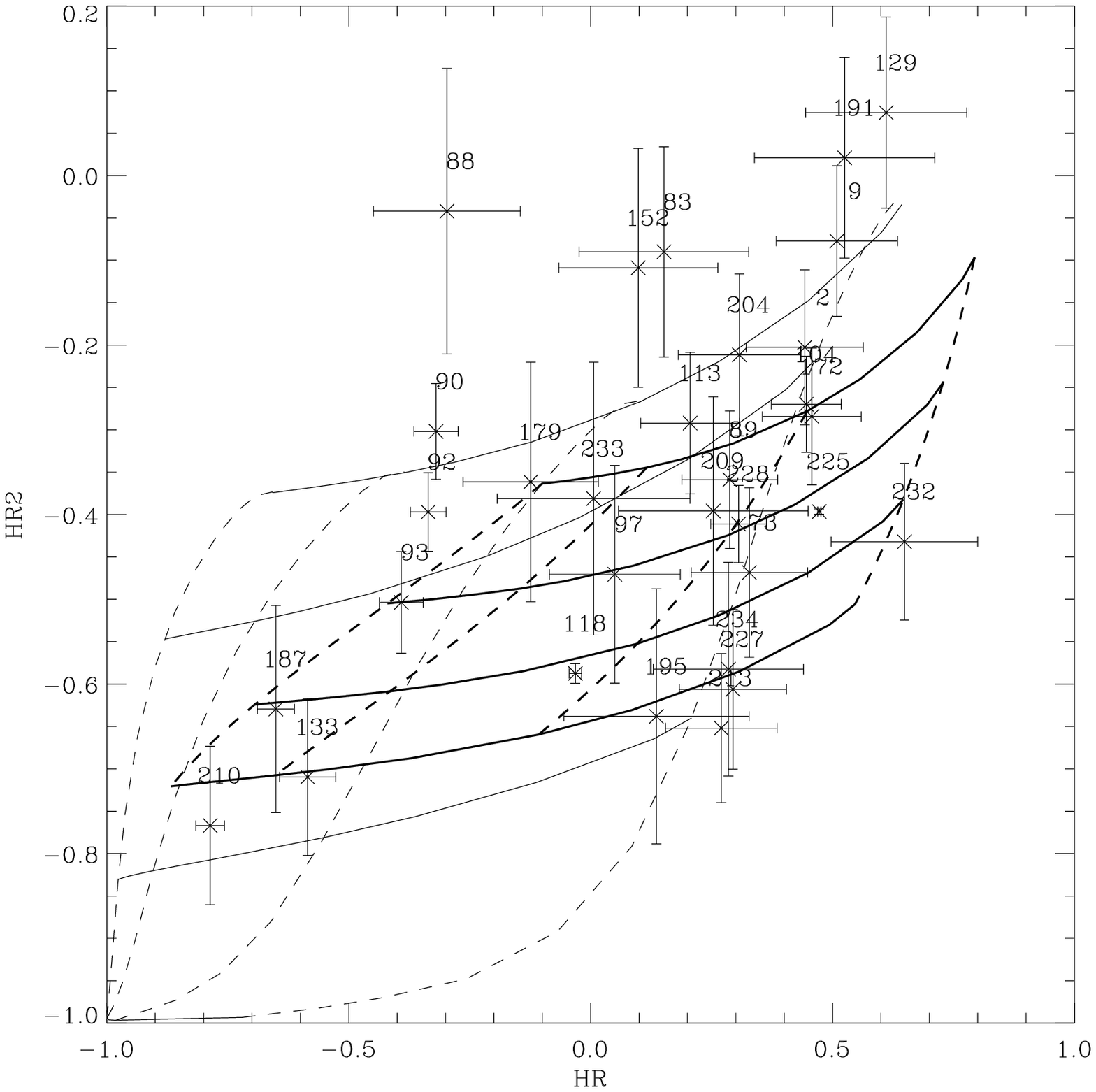,width=1\textwidth,angle=90, clip=}
}
\caption{\small Color-color diagrams of X-ray sources with the plotted
hardness ratios (HR and
HR2) and their 1$\sigma$  error bars as listed in Table~\ref{t:sou}.
The generic source numbers (Table~\ref{t:sou}) are marked. Also included in the plot are
hardness-ratio models: the solid thick curves are for the power-law model
with the photon index equal to 3, 2, 1, and 0, whereas the
solid thin curves are for a thermal plasma ({\small XSPEC} {\sl MEKAL}; $2\times$solar
metal abundances) with the temperature
equal to 0.3, 1, 3, and 6 keV (all from the bottom to the top). The absorbing
gas column densities are 1, 3, 10, and 30
$\times 10^{22} {\rm~cm^{-2}}$ for both models (dashed curves from the left to the right).
}
\label{f:hr}
\end{figure*}

The X-ray sources in the close vicinities of the \xs\ and \xq\
clusters are marked in Fig.~\ref{f:im_multi_bw}. Those in the
NICMOS fields are listed in Table~\ref{t:sou_cl}, including
apparent near-IR and radio counterparts, which most likely
represent massive stars if at the distance of the GC. In
particular, two of the three strongest X-ray sources in the \xs\
cluster have radio counterparts (AR1 and AR4; \citealt{lan05}). The faint and apparently resolved J174549.73-284926.1 is
probably associated with a close pair of radio sources of AR6 and
AR10 \citep{lan05}; AR6, in particular, has a unique nonthermal radio spectrum with
a spectral index of $-$0.7, whereas other radio sources in the field all have
positive indexes \citep{lan01b,lan05}.

To characterize the X-ray source number-flux relation (NFR) in the region,
we need to account for various complications involved in the
source detection and the confusion with interlopers (foreground
stars and background AGNs). For simplicity, our NFR analysis here uses only
the sources best-detected in the B band and with $P \le 10^{-7}$
(Table~\ref{t:sou}) and with count rates smaller than 
$ 2 \times 10^{-2} {\rm~counts~s^{-1}}$ (hence both LMXB 1E 1743.1-2843
and the transient XMMU J174554.4-285456 are excluded). This filtering, 
resulting in a sample of total 186 sources, minimizes the confusion with
foreground stars (typically with soft X-ray spectra and relatively low
sight-line absorptions) and background AGNs
(hard spectra and high absorptions), which should be preferentially detected
in either the 1-4 keV band (a total of 18 sources) or
the 4-9 keV band (29 sources), respectively.
Eleven of the 18  soft sources and nine of the 29 hard sources
are also detected in the 1-9 keV band, though not preferentially.
As will be shown in \S~\ref{ss:arches} and \S~\ref{ss:quin},
the absorptions towards the \xs\ and \xq\
are $\sim 5$ and $8 \times 10^{22} {\rm~cm^{-2}}$, sampling a
reasonable range of the column density toward the GC over the ACIS-I field
(Fig.~\ref{f:ob_rgb}). The average total column density through
the entire Galactic disk in the field is thus likely to be $\sim (1 - 2) \times 10^{23}
{\rm~cm^{-2}}$.

\begin{figure}[!tbh]
\centerline{
\psfig{figure=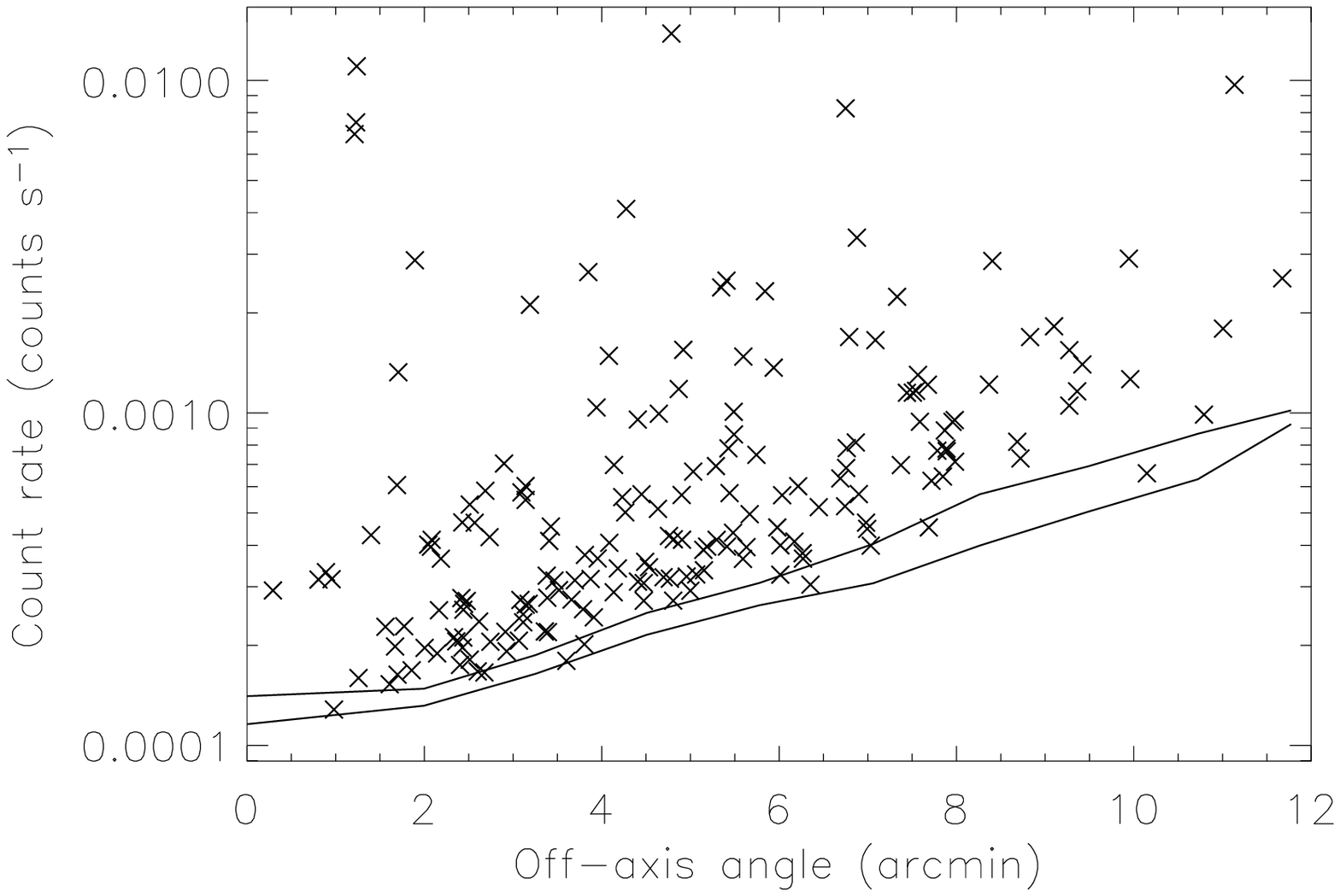,width=0.5\textwidth,angle=0,clip=}
}
\caption{\protect\footnotesize
Count rates of the sources best-detected in the 1-9 keV band versus their
off-axis angles in the 100 ks ACIS-I observation.
The curves illustrate the detection thresholds ($S_{min}$):
The upper curve is calculated via an azimuthal average, whereas
the lower curve is obtained by choosing the lowest value
in each concentric annulus around the aiming point of the observation.
}
\label{f:sou_limit_oa}
\end{figure}

The source detection completeness varies across the ACIS-I field,
depending on the local PSF, effective exposure, and background
\citep{wan04}. Fig.~\ref{f:sou_limit_oa} presents the dependence of both the
source count rate distribution and the detection completeness on
the off-axis distance. The detection limit of the count rate
varies from $\sim 1 \times 10^{-4} {\rm~counts~s^{-1}}$ near the
telescope axis to $10^{-3} {\rm~counts~s^{-1}}$ at the ACIS-I
corners. The detection is also subject to the so-called X-ray
Eddington bias: more intrinsically faint sources statistically
appear to have higher fluxes than the other way around
\citep{wan04}. We correct for both the incompleteness
and bias in our NFR analysis, following the approach detailed
in \citet{wan04}. Briefly, the NFR is analyzed as if it is an X-ray
spectrum with the field-integrated incompleteness and flux bias
included in the weighted effective area and response matrix.

We first estimate the background AGN contribution in our source
detection. We adopt the AGN NFR from
the \chandra\ deep surveys in the 2-10 band \citep{mor03}.
The energy flux in the NFR is converted into the 1-9 keV band
count rate, using the same intrinsic power-law spectrum with
$\Gamma = 1.4$ as assumed in \citet{mor03} and the sight-line
absorption $N_H = (1 - 2) \times 10^{23} {\rm~cm^{-2}}$ in our field.
Accounting for both the incompleteness and bias, we estimate the corresponding expected number
of AGNs  in the field to be 18-7,  consistent with nine
sources preferentially detected in the 4-9 keV band and excluded
from our NFR analysis of the sources that are best-detected in the 1-9 keV band.

\begin{figure}[htb!]
\centerline{
\psfig{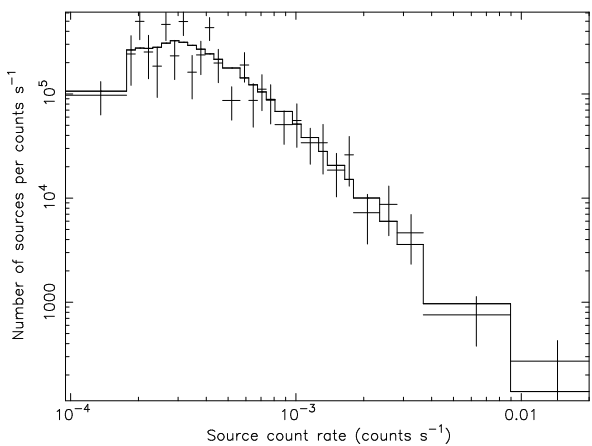}
}
\caption{\protect\footnotesize
Observed differential NFR of the sources
best-detected in the 1-9 keV band, compared with the best-fit
power-law model. The data are grouped to have a minimum four
sources per bin; the fit uses the Cash-statistic and is satisfactory,
judged from simulations in XSPEC.
}
\label{f:nfr}
\end{figure}

Fig.~\ref{f:nfr} shows the differential NFR of the all selected
sources (as in Fig.~\ref{f:sou_limit_oa}) and the best-fit power-law
\begin{equation}
\left({dN \over dS}\right)= A S^{-\alpha-1},
\label{dnds}
\end{equation}
where $S$ is in units of ${\rm~counts~s^{-1}}$, while $\alpha =
1.28_{-0.13}^{+0.14}$ and $A = 10^{-2.1_{-0.4}^{+0.5}}$ sources
per ${\rm~counts~s^{-1}}$ (error bars are all at the 90\% confidence).
To compare with similar results obtained by
\citet{mun06} for various GC regions, we adopt the same fiducial power
law spectrum with $\Gamma=1.5$ and absorbed by $N_H = 6 \times
10^{22} {\rm~cm^{-2}}$. The conversion from the 1-9 keV
count rate to the observed 0.5-8 keV photon flux is then $3 \times
10^{-3}$ ${\rm~(ph~cm^{-2}~s^{-1}})/({\rm counts~s^{-1}})$, while the
conversion to the corresponding absorption-corrected flux is $6 \times
10^{-11}$ ${\rm~(ergs~cm^{-2}~s^{-1}})/({\rm counts~s^{-1}})$. Thus
the detection limit of $\sim 1 \times 10^{-4} {\rm~counts~s^{-1}}$
corresponds to a 0.5-8 keV luminosity of $4 \times 10^{31}
{\rm~ergs~s^{-1}}$ at the distance of the GC.
To get the accumulated NFR, we convert $S$ to the above
photon flux, integrate
Eq.~\ref{dnds} to infinity, and account for our source
detection area of 278 arcmin$^2$. The resultant accumulated NFR is
\begin{equation}
N(<S) = N_0 \left(S \over S_0\right)^{-\alpha},
\label{ands}
\end{equation}
where the scaling factor $S_0 = 3 \times 10^{-6}
{\rm~ph~cm^{-2}~s^{-1}}$, and $N_0 = 0.14
{\rm~sources~arcmin^{-2}}$. The above parameter values can be compared to the
results based on 28 X-ray sources detected in
a 50 ks ACIS-I observation of the radio Arc region (\citealt{mun06}, their Table
2): $N_0 = 0.17 {\rm~sources~arcmin^{-2}}$ and $\alpha =
1.1\pm0.2$ (1$\sigma$ error bar), which largely overlaps in field with
the present observation. The two analyzes are in good
agreement. The slightly low $N_0$ value in the present analysis is
apparently due to our exclusion of both the very soft and hard sources (a
factor of 20\%). A re-analysis with these sources included
confirms this conclusion, but does not change the $\alpha$ value
significantly. This insensitivity to the exclusion of the
potential foreground and background interlopers indicates that the
above estimated NFR is robust.

\subsection{Arches Cluster}\label{ss:arches}

\subsubsection{Discrete X-ray Sources}\label{sss:sou_a}

\begin{figure}[!tbh] 
  \centerline{
      \epsfig{figure=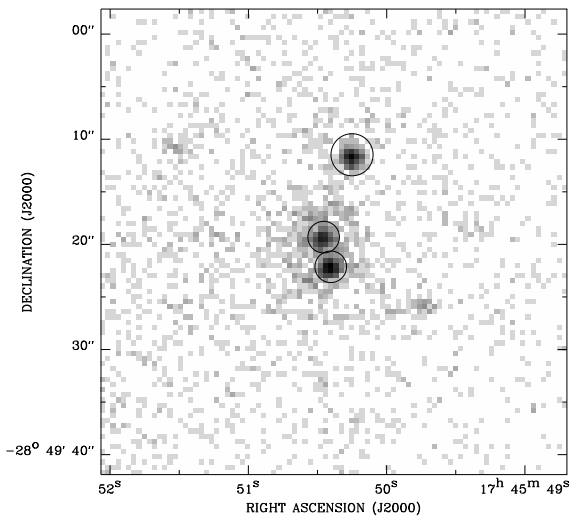,width=0.5\textwidth,angle=90}
    }
\caption{\small ACIS-I close-up of the \xs\ cluster core. The count image
is smoothed with a Gaussian with a FWHM of 0\farcs3. The circles
outlines the regions for the source spectral extractions.
Background is extracted within a concentric circle of 50\as\ radius, excluding
the source regions.} \label{f:core}
\end{figure}

\begin{figure*}[!hbt] 
  \centerline{
      \epsfig{figure=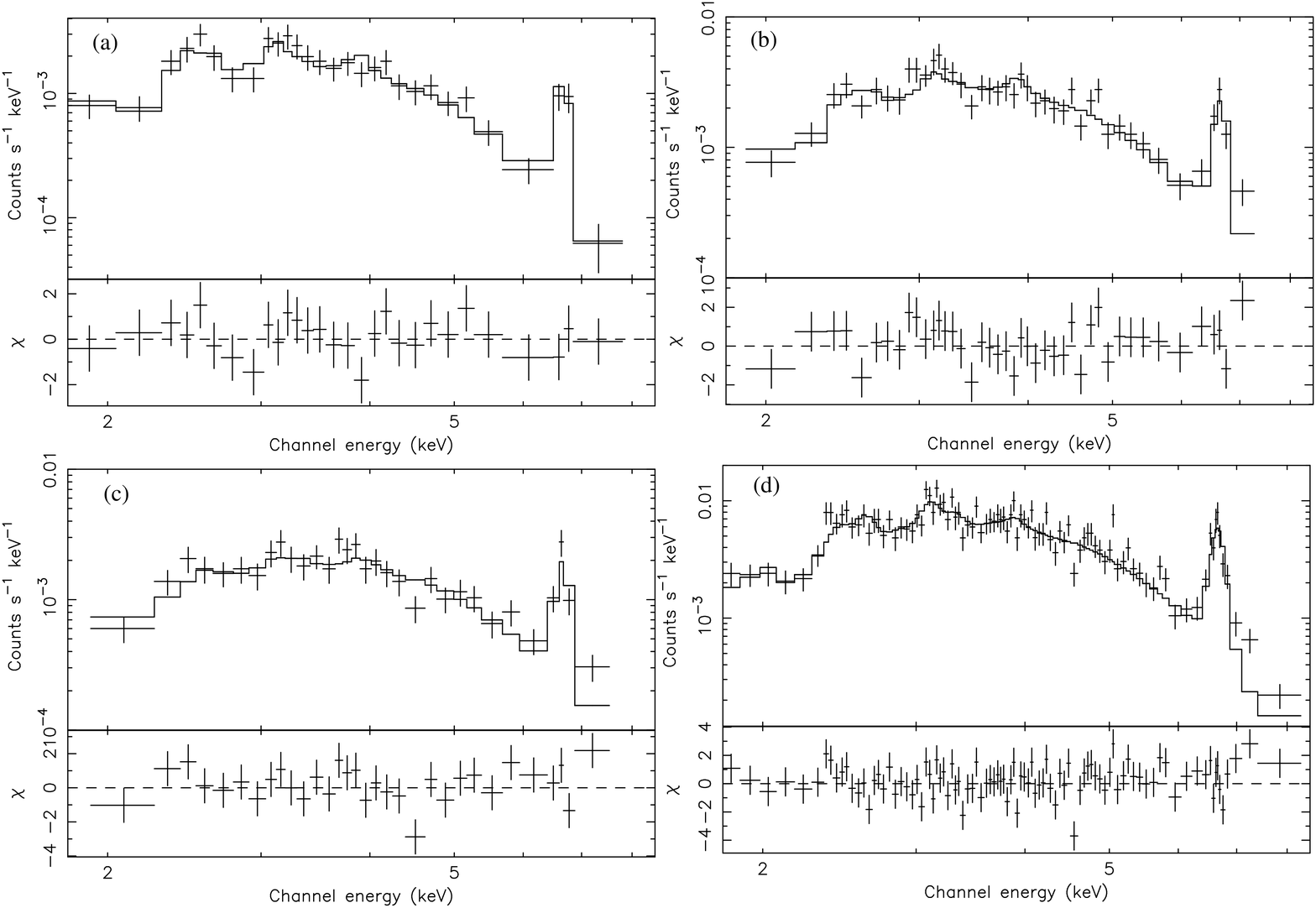,width=0.8\textwidth,angle=0}
    }
\caption{\small ACIS-I spectra of the three brightest X-ray sources in
the Arches cluster and the best-fit thermal plasma models: (a)
A1N, (b) A1S, (c) A2, and (d) the combination of A1N + A1S + A2.
The lower panels show the respective fit residuals relative to the 
errors of each bin. The spectra
are grouped to have at least 20 counts per  bin.} \label{f:spec_a}
\end{figure*}

The X-ray sources, A1N, A1S and A2, stand out in the \xs\ field
(Fig.~\ref{f:core}). The reasonably good counting statistics of
these sources allow for individual spectral analysis. We extract
the on-source 
spectra from the circle 
around each source
as illustrated in the figure. The spectra of the sources are
remarkably similar, in terms of both the overall spectral shape
and the presence of the strong 6.7-keV emission line
(Fig.~\ref{f:spec_a}). A
characterization of the spectra with an optically-thin thermal
plasma (XSPEC {\sl MEKAL} model) gives statistically consistent
temperatures and metal abundances as well as the foreground
absorptions (Table~\ref{t:spec_arches_p}), although there are significant
flux excesses above the model at $\gtrsim 7$ keV,
indicating the presence of a harder component (Fig.~\ref{f:spec_a}).

To further tighten the constraints on the intrinsic spectral
shape, we jointly fit the spectra of the three sources, reasonably assuming
that they have the same abundance and absorption, as members of
the same stellar cluster. The fit is satisfactory ($\chi^2/d.o.f.
=96/95$; Fig.~\ref{f:spec_a}d) and gives the best-fit parameters as
$N_{H}$=7.7$^{+0.8}_{-0.8}\times10^{22}$ ${\rm~cm^{-2}}$,
abundance=$1.8^{+0.8}_{-0.2}$ solar, and $kT = 1.8^{+0.2}_{-0.2}$
keV (A1N), $2.2^{+0.4}_{-0.3}$ keV (A1S) and $2.5^{+0.4}_{-0.3}$
keV (A2). 

\begin{deluxetable}{lccc}
  \tablecaption{Identifications of the X-ray Sources\label{t:sou_cl}}
  \tablewidth{0pt}
  \tablehead{
  \colhead{Source} &
  \colhead{Source$\tablenotemark{a}$} &
  \colhead{NIR$\tablenotemark{b}$}&
  \colhead{Radio$\tablenotemark{c}$} \\
  \noalign{\smallskip}
  \colhead{(1)} &
  \colhead{(2)} &
  \colhead{(3)} &
  \colhead{(4)} 
  }
  \startdata
\multispan{4}{\hfill Arches Cluster\hfill }\\
\\
85  &  A6 &2 (WNL)&AR6+AR10\\
90  &  A2 &9 (WNL)&--\\
92  &  A1S&6 (WNL)&AR1\\
93  &  A1N&7 (WNL)&AR4\\
\\
\multispan{4}{\hfill \xq\ Cluster\hfill }\\
\\
 214 &  QX1&242& -- \\
 215 &  QX5&231 (DWCL)&QR7\\
 216 &  QX2&257 (B0 I)&QR6\\
 217 &  QX3&211 (DWCL)&--\\
 219 &  QX4&--&-- \\
 244 &     &344 (B1 I-B3 I)&--  \\
\enddata
\tablenotetext{a}{Alternative X-ray source names given by
\citet{yus02,law04}.}
\tablenotetext{b}{Near-infrared counterparts: FMS1999 from
\citet{fig99a} for the Quintuplet and FNG2002 from \citet{fig02}
for the Arches: WNL - late-type WN stars (WN7-WN9); DWCL
- dusty late-type WC stars.} \tablenotetext{c}{Radio counterparts
from \citet{lan05}.}
\end{deluxetable}

\begin{deluxetable}{lcccccc}
\tabletypesize{\small}
\tablecolumns{7} \tablecaption{Spectral Fits
for X-ray Sources in the Arches Cluster} \tablewidth{0pt}
\tablehead{Name&$N_{H}$($10^{22}$ ${\rm~cm^{-2}}$)&kT
(keV)&Abundance&$\chi^2$/d.o.f&$L_{X}\tablenotemark{a}$&log($L_{x}$/$L_{bol}$)
 }
\startdata
A1N&$7.3^{+1.5}_{-1.1}$&$1.87^{+0.39}_{-0.32}$&$2.8^{+10.1}_{-1.5}$&16.9/23&7.2&-5.8\cr
A1S&$8.1^{+1.1}_{-1.2}$&$2.1^{+0.58}_{-0.34}$&$1.5^{+1.2}_{-0.6}$&42.2/40&11&-5.7\cr
A2&$6.4^{+2.5}_{-1.6}$&$3.25^{+2.62}_{-1.24}$&$1.6^{+2.1}_{-0.6}$&33.9/28&4.6&-5.9\cr
\enddata
\tablenotetext{a}{The luminosity is in units of $10^{33} {\rm~ergs~s^{-1}}$
and in the 0.3-8~keV band.}
\label{t:spec_arches_p}
\end{deluxetable}

The spectra shown in Fig.~\ref{f:spec_a} represent a substantial
improvement in quality than those in \citet{yus02}. The on-axis
spatial resolution of our new observation
allows us not only to separate the spectra of A1N and A1S, but
also to minimize the contamination of surrounding diffuse
emission (Fig.~\ref{f:core}). The calibration of the data has also been
improved significantly (e.g., the inclusion of the CTI
correction). These improvements probably account for the
discrepancies between the present results and those presented in
\citet{yus02}. Our analysis shows that one-temperature
plasma is adequate to fit each of the above spectra and that our
inferred total 0.2-10 keV luminosities of the sources are smaller than that of
\citet{yus02}  by a factor more than 15. These two
discrepancies are actually related. The use of the two-temperature
plasma model in \citet{yus02} required a very
high hydrogen column density
($N_{H}=12.4^{+2.9}_{-2.0}\times10^{22} {\rm~cm^{-2}}$), which in
turn gave a large absorption-corrected luminosities.

\subsubsection{Diffuse X-ray Emission}\label{sss:dif_a}

\begin{figure*}[!tbh]
  \centerline{
      \epsfig{figure=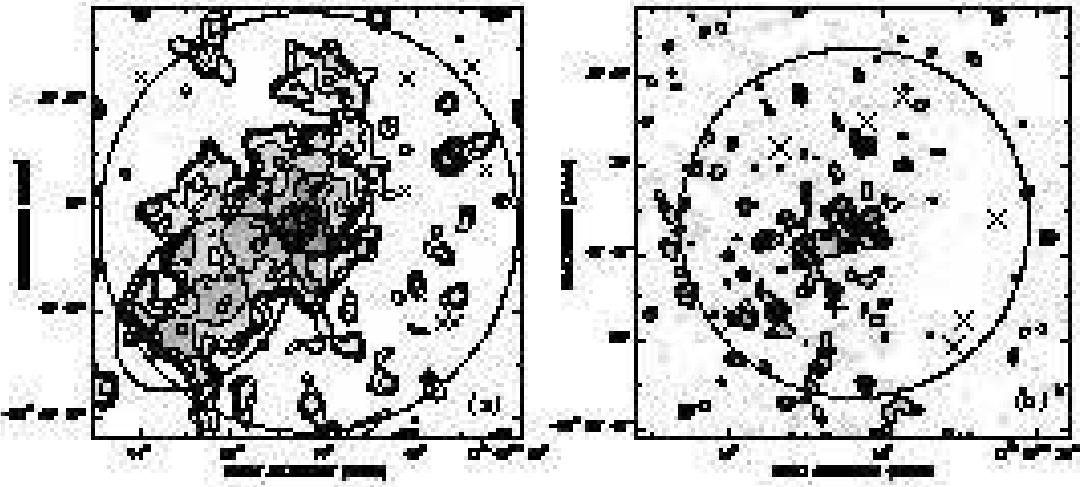,width=0.9\textwidth,angle=0}
    }
\caption{\small Diffuse X-ray emission intensity distributions of the
\xs\ (a) and \xq\ (b) regions. The excised source positions are marked
(see Fig.~\ref{f:im_multi_bw}). These X-ray images are adaptively
smoothed to achieve a background-subtracted
signal-to-noise ratio of $\sim 6$ to show low-surface brightness
emission. The intensity contour levels are at 2.1, 2.3, 2.9, 4.1,
6.4, 11, 21, and 39 (above a local background of 1.8) for (a) and at
3.3, 4.1, and 5.7 (above a local background of $\sim 2.5$) for (b); all in units of $10^{-3}
{\rm~counts~s^{-1}~arcmin^{-2}}$. The large circles in (a) and (b) 
outline the regions that we use to estimate the total diffuse
X-ray emission from the clusters. The two ellipses in (a) outline
the regions from which the spectra in Fig.~\ref{f:spec_dif_a} are
extracted. The {\sl plus} signs mark the centroid positions of the clusters. }
\label{f:im_dif}
\end{figure*}

Fig.~\ref{f:im_multi_bw} and \ref{f:im_dif} show that the enhanced
diffuse X-ray emission is distributed over a region greater than
the stellar core of the \xs\ cluster. The diffuse X-ray
enhancement is quite isolated within a radius $r \sim 60$\as\  and
is spectrally harder than large-scale diffuse X-ray emission  in
the region to the southeast (Fig.~\ref{f:ob_rgb}). We extract a
spectrum of this enhancement from this radius and a background
from a 100\as\ circle to the west  within the same
CCD chip. The background-subtracted spectrum of the diffuse emission
exhibits significant line emission in the energy range of 6.4 to
6.7 keV. The continuum-subtracted narrow band images of the
diffuse emission (\S~\ref{ss:chandra}) further show that the $\sim
6.7$-keV line emission arises in a plume from the cluster core
(Fig.~\ref{f:im_line}).
This plume has a size of $\sim 30$\as\ at the \xs\ cluster and elongated toward the northeast.
The 6.4-keV line emission is certainly more extended, although
its exact extent is difficult to determine; low surface
brightness 6.4-keV line emission of similar EW is ubiquitous in
the GC \citep{wan02a}. This enhancement
of the 6.4-keV line (as well as the continuum emission) around the \xs\
cluster is particularly strong in an extension from the cluster toward the southeast (SE).
The overall morphological appearance of this enhancement is quite
irregular (see \S~\ref{ss:cs_dist} for a discussion on the possible connection to
the adjacent ``$-$30 km s$^{-1}$'' molecular cloud).

\begin{figure*}[bht,angle=0]
  \centerline{
      \epsfig{figure=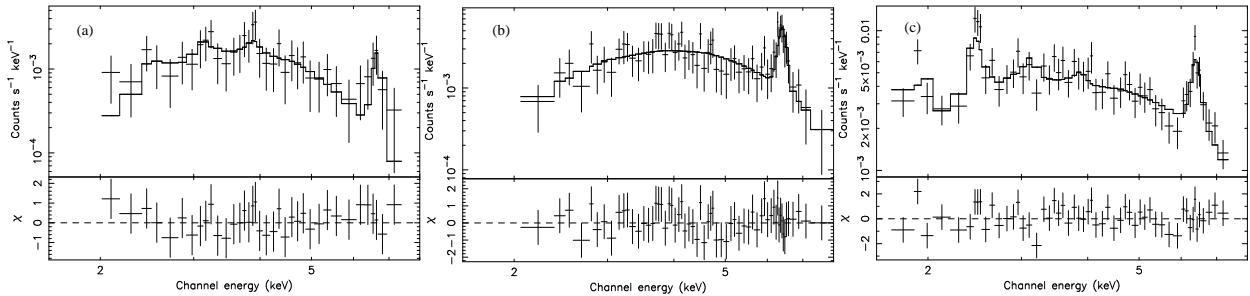,width=1.\textwidth,angle=0}
    }
\caption{\small ACIS-I diffuse X-ray emission spectra: the 6.7-keV
plume (a), the southeast extension (b), and the low-surface
brightness outer region (c) of the \xs\ cluster. The spectra,
grouped with a background-subtracted S/N$>$3, are fitted with the
{\sl NEI} for (a), {\sl PL+GAU} for (b) and {\sl
MEKAL+PL+GAU} for (c) (Table~\ref{spec_d}).}
 \label{f:spec_dif_a}
\end{figure*}

\begin{deluxetable}{lccccc}
\tablecolumns{6} \tablecaption{Diffuse X-ray
spectral fits for the Arches
cluster}\tablewidth{0pt} \tablehead{Region&Model&
Key Model Parameters &$N_{H}$&$\chi^2$/d.o.f & $L_X$
 }
\startdata
Central Plume &{\sl NEI}   &$kT =2.56 (> 1.18)$, $\tau =10^{1.1 (> 0.9)}$ &$11.0^{+4.0}_{-2.4}$ &12.0/33 &3.8\\ 
              &{\sl MEKAL} &$kT =1.88^{+4.19}_{-0.61}$ &$10.8^{+5.3}_{-5.6}$ &13.3/34 &3.2\\
SE Extension  &{\sl PL+GAU} &$\Gamma=1.3^{+1.4}_{-1.1}$ &$ 6.2^{+2.7}_{-5.6}$&29.2/57 &4.1\\
LSBXE         &{\sl MEKAL+POW+GAU} &$kT=0.45_{-0.10}^{+0.25}$, $\Gamma=1.3$ (fixed) &$9.2^{+1.8}_{-2.3}$& 62.3/52 & 12\\ 
\enddata
\tablecomments{The spectral model names are from XSPEC: {\sl NEI} -
non-equilibrium ionization collisional plasma; {\sl MEKAL} - collisional ionization
equilibrium plasma; {\sl PL} - power law; and {\sl GAU} - Gaussian line. The
metal abundances of plasma is fixed to be 2$\times$solar, as
inferred from the point-like source spectra. The plasma
temperature ($kT$), ionization time scale  ($\tau$), absorption
column density ($N_H$),
and the 2-8 keV luminosity ($L_X$) are in the units of
keV, ${\rm~cm^{-3}~s}$, $10^{22} {\rm~cm^{-2}}$,
and $10^{33} {\rm~ergs~^{-1}}$,
respectively.} \label{spec_d}
\end{deluxetable}

\begin{figure}[!hbt]
  \centerline{
      \epsfig{figure=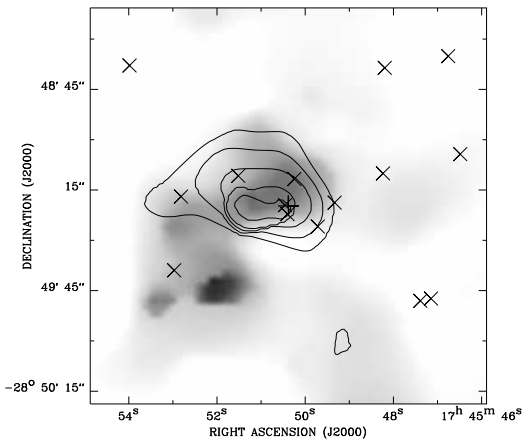,width=0.4\textwidth,angle=90}
    }
\caption{\small ACIS-I 6.4-keV and 6.7-keV line intensity maps of the \xs\ cluster region.
The gray-scale (6.4-keV intensity) is in the range of 0.5 to 4
$\times 10^{-3} {\rm~counts~s^{-1}~arcmin^{-2}}$ (see also Fig.~\ref{f:cs_x}),
while the overlaid 6.7-keV line intensity contours are
at 0.6, 0.8, 1, 1.2, and 1.6 $\times 10^{-3} {\rm~counts~s^{-1}~arcmin^{-2}}$.
The {\sl plus} sign marks the centroid position of the
cluster. }
    \label{f:im_line}
\end{figure}

We extract two diffuse X-ray spectra: one from the central 6.7-keV line plume,
and the other from the SE 6.4-keV line extension from the regions
outlined in Fig.~\ref{f:im_dif}(a). We fit the spectra with
a power law plus a Gaussian line (its width is fixed to zero)
to characterize the Fe line centroid and EW, which are $6.60^{+0.10}_{-0.14}$ keV and
$1.2^{+1.0}_{-0.9}$ keV for the plume and
$6.39^{+0.05}_{-0.05}$ keV and $1.4^{+0.9}_{-0.5}$ keV for the SE extension, respectively. The line and continuum fluxes are 0.25 and 1.3 for the
plume, and 0.64 and 2.8 for the SE extension;
all in units of $10^{-13} {\rm~ergs~s^{-1}~cm^{-2}}$.
The fits are all satisfactory.
But the power law component is reasonably constrained only for the SE extension, and the
fitted parameters are included in Table~\ref{spec_d}. The best-fit power
law index and line centroid of the SE extension spectrum
agree well with the theoretical prediction for the emission
from low-energy cosmic-ray electrons interacting with the
ambient medium (\citealt{val00}; see \S~\ref{ss:d_line}).

While a collisionally ionization equilibrium plasma gives
a good fit to the spectrum of the plume (Table~\ref{spec_d}),
the 6.4-keV line in the spectrum of the SE extension may indicate
a plasma in a non-equilibrium ionization
(NEI) state (see \S~\ref{s:dis}). We thus try a fit of the spectrum with
the {\small XSPEC} {\sl NEI} model with a metal abundance equal to 2$\times$solar.
The fit to the spectrum of the SE extension requires an ionization time scale  of
$\tau \sim n_e t < 1.1 \times10^{10} {\rm~cm^{-3}~s}$, too small to be consistent
with any dynamic model of the plasma on the observed spatial scale  (\S~\ref{s:dis}).

We further extract a spectrum of the low-surface brightness X-ray
emission (LSBXE) from the large circle in Fig.~\ref{f:im_dif}a
minus the plume and the extension regions. This spectrum
(Fig.~\ref{f:spec_dif_a}c) also shows both the 6.4-keV line
and the He-like S XV K$\alpha$ line at $\sim 2.5$ keV,
indicating a mixture of multiple components. Motivated by the above spectral
analysis of the plume and the SE extension, we characterize the LSBXE spectrum,
using a simple combination of a {\sl MEKAL} plasma, a power law with 
$\Gamma = 1.3$, and a Gaussian line with its centroid fixed at 6.4 keV.
This combination gives a reasonable fit to the spectrum, and the fitted
parameters are included in Table~\ref{spec_d}.

\subsection{\xq\ Cluster}\label{ss:quin}

\subsubsection{Discrete Sources}\label{sss:sou_q}

\begin{figure*}[!bht]
  \centerline{
      \epsfig{figure=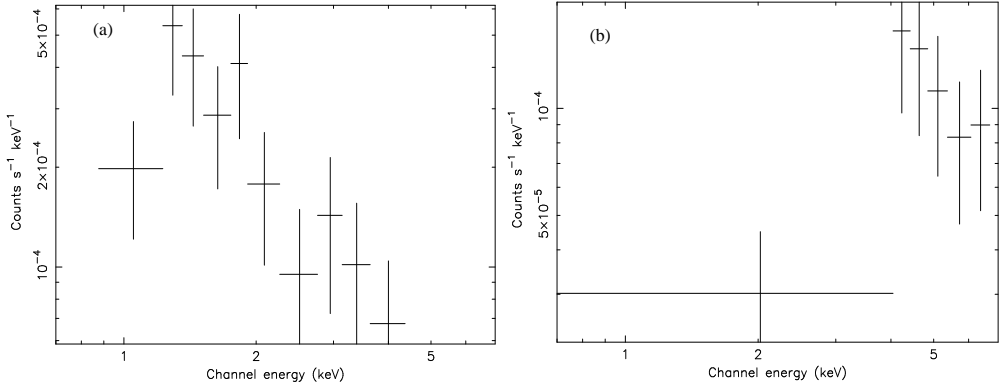,width=0.8\textwidth,angle=0}
    }
\caption{\small Example spectra of X-ray sources in the \xq\ core: QX1
(a) and QX4 (b).
} \label{f:spec_sou_q_ind}
\end{figure*}

\begin{figure}[htb,angle=-90]
  \centerline{
      \epsfig{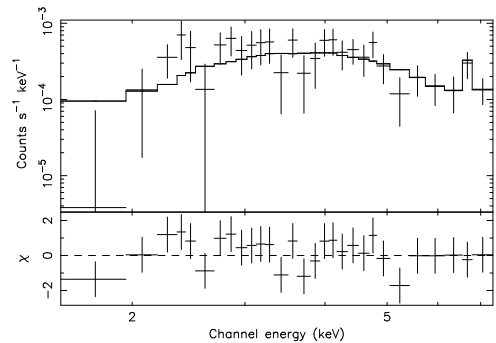}
    }
\caption{\small Combined ACIS-I spectrum of QX2-4 and the best-fit
thermal plasma model. } \label{f:spec_sou_q}
\end{figure}

Within the field of view of the NICMOS observation
(Fig.~\ref{f:im_multi_bw}d), we find eight X-ray sources
(Table~\ref{t:sou_cl}).  J174614.67-284940.3 is located close to a
source candidate first suspected by \citet{law04} (their QX5 or
J174614.7-284947, which should have been named J174614.7-284942). 
This faint source is now well separated from QX1
and has a near-IR counterpart [FNG2002] 231. Compared to those in
the Arches cluster, all of the eight sources are rather faint; in
particular, the total number of counts of the four relatively bright 
sources in the core (QX1-4) is only about $4 \times 10^2$. They also show diverse
spectral characteristics, as indicated by their significantly
different hardness ratios (Table~\ref{t:sou}).
Fig.~\ref{f:spec_sou_q_ind} presents two extreme examples of the
source spectra. It is clear that QX1 is very soft, whereas QX4
appears extremely hard. The spectral characteristics of QX2 and
QX3 fall between these two extremes. To quantify the diversity, we
assume that all these sources have an approximately same intrinsic
spectral shape, but have different foreground absorptions. A
joint fit of the spectra with a {\sl MEKAL} plasma model (assuming a
metal abundance of 2$\times$ solar) gives a characteristic
temperature of $>$8.52 keV and the absorptions along the
sight-lines to QX1, QX2, QX3, and QX4 as $N_{H}$($10^{22}
{\rm~cm^{-2}}) = 1.3^{+0.5}_{-0.3}$, $4.8^{+2.4}_{-1.3}$,
$4.3^{+2.4}_{-1.5}$, and $9.3^{+6.9}_{-3.5}$, respectively.
Clearly, the absorption toward QX1 is significantly smaller than
toward other sources. Thus QX1 is likely a foreground star.
Without a near-IR counterpart, QX4 is likely a background source
(e.g., an AGN) or a stellar object that is still deeply embedded
in dense gas. QX2 and QX3 do have near-IR counterparts tentatively
classified as B0I and dust-enshrouded WCL stars
(Table~\ref{t:sou_cl}). Because of this diversity, we cannot rule
out that QX4 is a member of the \xq\ cluster. We thus fit the
combined spectrum of QX2, QX3 and QX4, which have relatively
comparable spectral characteristics. The accumulated spectrum
shows an emission line at $\sim 6.7$ keV and can be characterized
($\chi^2/d.o.f. = 34.4/31$) by a {\sl MEKAL} model with $kT =
8.68^{+9.05}_{-3.99}$ keV and a foreground absorption of $N_{H} =
5.9^{+1.9}_{-1.3}$$\times10^{22}$ ${\rm~cm^{-2}}$
(Fig.~\ref{f:spec_sou_q}), which is consistent with $A_V =
29.0^{+5}_{-5}$ of this cluster (1$\sigma$ error bar;
\citealt{fig99a}), assuming $N_H/E(B-V) \approx 5 \times 10^{21}
{\rm~cm^{-2}~mag^{-1}}$ \citep{boh78} and $A_V/E(B-V) \approx
2.6-5.5$ \citep{sch98}. The total absorption-corrected
0.3-8 keV luminosity is $7.6\times10^{32}$ ${\rm~ergs~s^{-1}}$.

\subsubsection{Diffuse emission}\label{sss:dif_q}

\begin{figure}[htb,angle=-90]
  \centerline{
      \epsfig{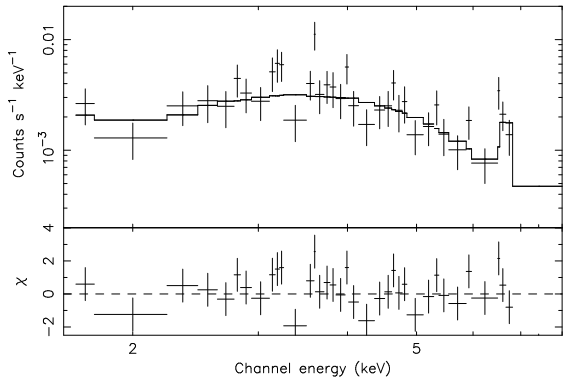}
    }
\caption{\small ACIS-I spectrum of the diffuse emission in
the \xq\ cluster and the best-fit thermal plasma model.}
\label{f:spec_dif_q}
\end{figure}

The extent of the diffuse X-ray enhancement around the \xq\
cluster is uncertain (Fig.~\ref{f:im_dif}). The cluster seems to
be embedded in a large-scale diffuse X-ray-emitting region,
although the spectrum of the enhancement appears to be
slightly harder than that of the surrounding region (Fig.~\ref{f:ob_rgb}).
We extract a spectrum of the diffuse emission from a circle of
$r =1^\prime$ radius around the \xq\ centroid
(Fig.~\ref{f:im_dif}) and a background spectrum from the field
within a concentric annulus of r=1$^\prime$-2$^\prime$. The
background-subtracted spectrum can be characterized
($\chi^2/d.o.f. =40.5/34$) by a {\sl MEKAL} plasma model (again assuming
a metal abundance of $2\times$solar) with $kT =10^{+4.6}_{-2.7}$ keV
and $N_H = 3.8^{+0.7}_{-0.5} \times10^{22}$ ${\rm~cm}^{-2}$
(Fig.~\ref{f:spec_dif_q}). These parameters are consistent  with
the values obtained from the fit to the combined spectrum of the
discrete sources in the core of the \xq\ cluster. The
absorption-corrected luminosity of the diffuse emission in the
2-8 keV range is $3\times10^{33} {\rm~ergs~s^{-1}}$. 

\subsection{Molecular Gas near the Arches Cluster}\label{ss:cs_dist}

\begin{figure*}[!htb]
  \centerline{
       \epsfig{figure=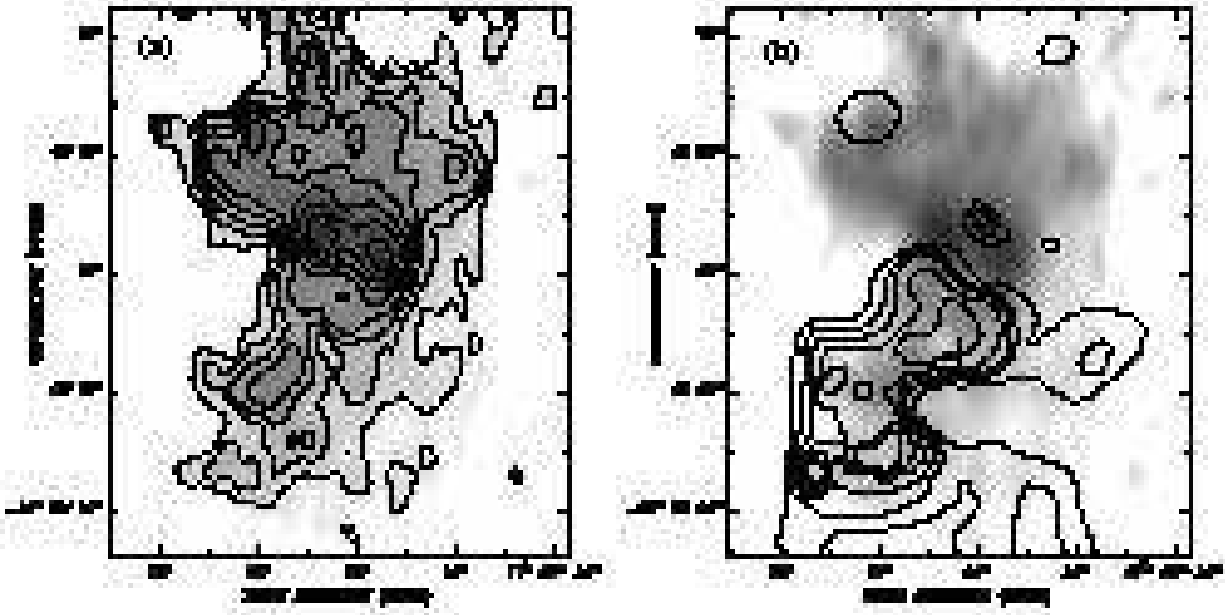,width=1.\textwidth,angle=0}
     }
  \caption{\small (a) Distribution of CS (J=2-1) emission integrated over the central channels,  with contour levels representing 9, 12, 15, 18, 21, 24,
27, and 30 ${\rm~Jy~beam^{-1}~km~s^{-1}}$. (b)  The same CS image overlaid with the 6.4-keV line intensity
contours at 0.7, 0.9, 1.1, 1.5, 2.1, and 2.9 $\times 10^{-3} {\rm~counts~s^{-1}~arcmin^{-2}}$.
The {\sl plus} sign marks the centroid position of the \xs\ cluster.
    \label{f:cs_x}}
\end{figure*}

Fig.~\ref{f:cs_x} presents a comparison of the distribution of the
CS (J=2--1) line emission and the 6.4 -keV line
emission (Fig.~\ref{f:im_line}). This ``clump'' of molecular gas represents
one of the easternmost parts of the ``$-$30 km s$^{-1}$ cloud'' (and corresponds
to Peak 2 in the single dish study of this molecular cloud;  \citealt{ser87}). 
The filamentary molecular cloud has an average velocity of
 $\sim -25 {\rm~km~s^{-1}}$, although there are large velocity gradients over
the cloud and the FWHM of the line is up to $\sim 30 {\rm~km~s^{-1}}$. 
In contrast,  the \xs\ cluster has an average velocity of $\sim +95 {\rm~km~s^{-1}}$
\citep{fig02}. Therefore, the relative velocity  between the cluster
and the cloud is at least  $v_r \sim 120 {\rm~km~s^{-1}}$.

Fig.~\ref{f:cs_x}a shows an image of the CS (J=2-1) emission integrated over the central channels, where the line emission is present (i.e., velocities of $-$5 to $-$40 km s$^{-1}$). Fig.~\ref{f:cs_x}b compares this image to the distribution of the diffuse 6.4 keV X-ray emission. There is little morphological similarity between the molecular line emission and the diffuse X-ray emission. The CS intensity is the
strongest in the north, where there is little X-ray emission
enhancement, either in the broad band or in the Fe K$\alpha$ lines. 
Although the 6.4-keV emission appears to coincide
spatially with the southern extension of the molecular gas, Fig.~\ref{f:cs_x}b shows 
little peak-to-peak correlation. The individual channel images also reflect this distribution and the lack of detailed physical correlation with the 6.4 keV X-ray emission. 

\section{Discussion}\label{s:dis}

The above results show distinctly different X-ray properties between the \xs\ and \xq\ clusters.
The \xs\ cluster contains three luminous point-like sources, all of which exhibit the strong 6.7-keV
emission line, and two apparently diffuse components with either 6.4-keV or 6.7-keV line emission. The 6.4-keV line-emitting enhancement is strongly elongated, morphologically, tracing
the east boundary of the CS cloud's southern extension.
These characteristics are absent in the \xq\
cluster, in which we detect only weak X-ray sources, plus a very low surface brightness
diffuse emission with a hard spectrum. There is also no evidence for any
associated CS cloud. In the following, we discuss origins of these
various X-ray components, possible causes of the distinct differences in the X-ray properties between
the two clusters, and implications of our results.

\subsection{Galactic Center Environment}\label{ss:d_envi}

We attempt to  understand the \xs\ and \xq\ clusters in the
context of the unique GC environment. The generally high gas density
and pressure, strong gravitational tidal force, and large random and
bulk motion  velocities in the GC affect both the formation and
evolution of young stellar clusters \citep{mor93}. Here we concentrate on the potential interplay between
the molecular gas and the \xs\ cluster.

Is the ``$-$30 km s$^{-1}$ cloud'' and the \xs\ cluster physically associated? On
one hand, because of their large velocity separation, the two
systems would pass each other in only $\sim 10^4$ yrs if the
 size of the cloud along the line of sight 
is comparable to that  projected in
the sky and at the GC distance. On the other hand, the volume
filling factor of dense molecular gas in the region is quite high
($\gtrsim 0.3$; \citealt{ser87}). In particular, the well-known
Arched filaments all have negative velocities similar to that of
the molecular gas; the photon-ionization modeling of these filaments
suggests that they are physically in the vicinity of the \xs\
cluster \citep{lan01a}. Thus the probability for a chance physical contact of a
dense cloud with the cluster is not small.
An independent argument for the association is an effective
extinction deficit of $\delta A_V \approx 10$ over a region of
$\sim 15$\as\ from the cluster core, which can be interpreted as the displacement of
the dusty gas by the cluster wind and/or the dust grain destruction by the 
UV radiation from the cluster
\citep{sto02}. The extinction is the largest towards the region just west of the cluster
(\citealt{sto02}; Note that East is to the right in their Figs.~3 and 8). Interestingly,
this extinction deficit, corresponding to $\delta N_H \sim 3
\times 10^{22} {\rm~cm^{-2}}$, provides a natural explanation for
the difference between our measured X-ray-absorbing column $N_H
\approx 8 \times 10^{22} {\rm~cm^{-2}}$ and the prediction from
the total sight-line extinction $A_V = 24$ \citep{sto02, boh78, sch98}. Furthermore, the
interaction of the \xs\ cluster wind with the cloud may also
explain the strong and distinct X-ray emission enhancement around
the \xs\ cluster (\S~\ref{ss:d_line}).
The far-IR spectroscopy further shows the presence of
a component of dusty gas at a velocity of $-70 
{\rm~km~s^{-1}}$, unique at the location of the \xs\ cluster
\citep{cot05}. This component
may represent shocked cloud gas, deflected toward us (e.g., in 
the lower left direction of Fig.~\ref{f:ill}; see \S~\ref{ss:d_line}
for further discussion). 
Therefore, we tentatively conclude that the
``$-$30 km s$^{-1}$ cloud'' and the cluster are undergoing a collision.

The collision of such clouds with the \xs\ cluster may have
strongly affected its evolution. The absence of a natal
cloud associated with the cluster at its velocity, for example,
may be a consequence of the collision. The removal of this natal
cloud from the cluster at an early time could have reduced the
probability for low-mass stars to form. The cloud-cloud collision could
also be responsible for the formation of the \xs\ cluster itself. The exceptionally
high gas temperature and velocity dispersion in such a formation formation process
could also result in a top-heavy initial mass function (IMF; see
\S~\ref{ss:d_yso} for further discussion).

Our X-ray study further provides useful measurements about the GC environment.
In addition to the $N_H$ measurement, we have also directly estimated
the metal abundance
(mainly iron) in the GC. Recent estimates based on near-IR
spectroscopy of young and intermediate-age supergiants in GC
\citep[e.g.,][]{ram00} suggest an iron abundance that is consistent
with being solar, i.e., similar to the abundance observed in the
solar neighborhood. This result is {\sl against} the general trend of an
increasing metallicity with decreasing galacto-centric radius
as observed in the disks of the Milky Way and nearby galaxies. Our X-ray
measured iron abundance of $\sim 1.8^{+0.8}_{-0.2}$ solar,
based on the spectral analysis of the luminous colliding wind candidates
in the \xs\ cluster, agrees
with the trend. The thermal process involved in the X-ray emission
is quite simple, and the ion fraction of the He-like Fe K$\alpha$
emission is insensitive to the exact plasma temperature fitted.
Furthermore, the iron abundance in the winds of the massive stars
is not expected to be contaminated by their own nuclear synthesis
in the deep cores of the stars. Therefore, we conclude that the
iron abundance  in the ISM of the GC is super-solar.

\subsection{Nature of Discrete X-ray Sources}\label{ss:d_sou}

As shown in \S~\ref{ss:sou_pop}, our analysis confirms a relatively flat
source NFR in the region of the \xs\ and \xq\
clusters, as indicated first in \citet{mun06}.
Our obtained power law slope ($\alpha = 1.26_{-0.13}^{+0.14}$;
90\% confidence) is flatter than those in the deep observations
of Sgr B2 ($1.7\pm0.2$) and Sgr A$^*$ ($1.4\pm0.1$) as well
as the $2^\circ \times
0\fdg8$ shallow survey ($1.5\pm0.1$). The implied over-population
of relatively bright X-ray sources is clearly related to the
presence of the two clusters.

The discrete X-ray sources in the core of the clusters are
unlikely due to emission from individual normal massive stars or
even binaries. The X-ray emission from such a star/binary can be
characterized typically  by an optically-thin thermal plasma with
a temperature of $\sim 0.6$~keV and a luminosity following the
empirical relation $\frac{L_{X}}{L_{bol}} \sim 10^{-7}$, where
$L_{bol}$ is the bolometric luminosity. Thus the emission is too
soft and faint to be observed from the GC. Even the Pistol star
near the core of the \xq\ cluser (Fig.~\ref{f:im_multi_bw}d) is
not detected as an X-ray source. The star is a luminous blue
variable with $L_{bol} \gtrsim 10^{6.6} L_\odot$ and has an
extinction of $A_{K} \approx 3.2$, corresponding to $N_{H} \approx
5.1\times10^{22} {\rm~cm^{-2}}$. Assuming the {\sl MEKAL} thermal
plasma with a temperature of 0.6 keV, we estimate that the 3$\sigma$ upper limit to the
0.3-8 keV luminosity is 3 $\times10^{33} {\rm~ergs~s^{-1}}$,
consistent with $L_{x}/L_{bol} \sim 10^{-7}$.

Most likely, the luminous X-ray sources associated with the \xs\
cluster represent colliding stellar winds in massive star close binaries.
The characteristic shock temperature of a colliding wind is
\begin{equation}
T \simeq (3\times10^{7} {\rm~K}) v_{w,3}^{2},
\end{equation}
where $v_{w,3}$ is the relative
 colliding wind velocity in units of $10^3 {\rm~km~s^{-1}}$.
Well-known examples of such systems are WR11 (kT$\approx
4.3$ keV, $L_X \sim 8 \times 10^{33} {\rm~ergs~s^{-1}}$;
\citealt{sch04}) and WR140 (kT$\approx 3$ keV, $L_X
\sim 2 \times 10^{33} {\rm~ergs~s^{-1}}$; \citealt{zhe00}). Clearly, the expected temperatures are similar to the
measured values for the sources in the \xs\ cluster, although their
luminosities seem to be substantially higher than those confirmed
colliding wind systems, which all have $L_X < 1 \times 10^{34}
{\rm~ergs~s^{-1}}$ \citep[e.g.,][]{osk05}. The unusually high X-ray luminosities of the
colliding wind systems may be related to the compactness of
the \xs\ cluster, in which very close binaries may form dynamically.

In contrast, the X-ray sources in the \xq\ cluster are probably
typical colliding wind systems. They all have  individual
$L_X$ in the range of $(0.2-3) \times 10^{33}
{\rm~ergs~s^{-1}}$  as well as the hard X-ray
spectra with the 6.7-keV emission line, as expected.

\begin{figure*}[!htb]
  \centerline{
      \epsfig{figure=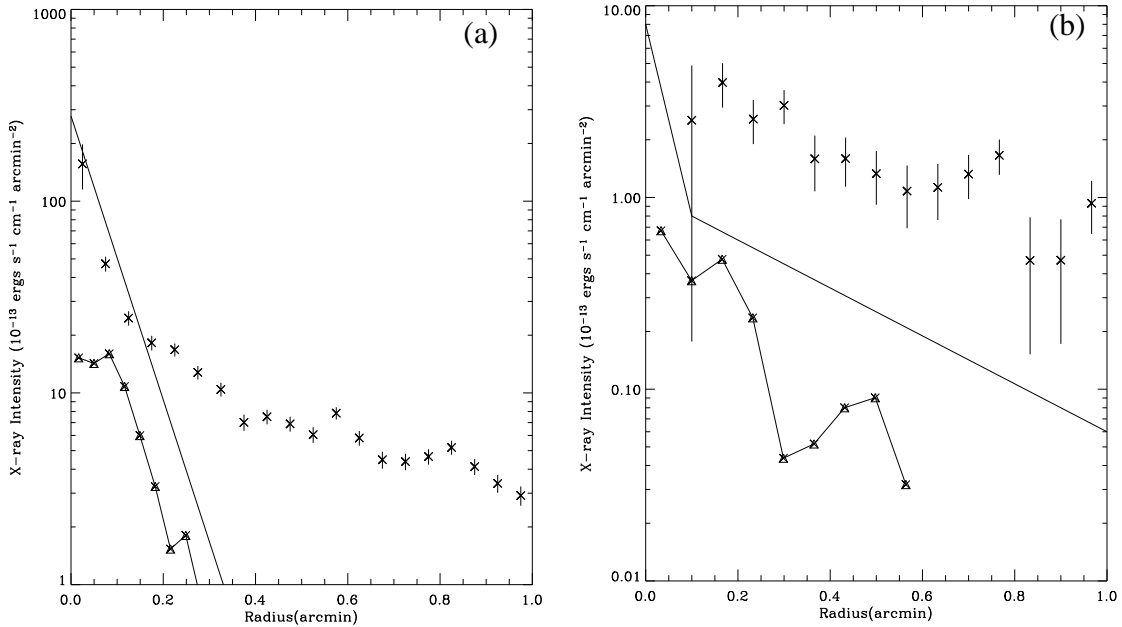,width=1\textwidth,angle=0}
     }
  \caption{\small Radial ACIS-I 1-9 keV intensity profiles ({\sl crosses} with $1\sigma$ error
bars) around the \xs\ (a) and \xq\ (b) clusters, compared with the
respective NICMOS F205W stellar light distributions (connected
{\sl triangles}). The cluster wind predictions are shown
approximately as the solid line
from 3-D simulations for the ``standard'' stellar wind mass-loss rates of the
two clusters \citep{roc05}.
    \label{f:rbp}}
\end{figure*}

While only relatively luminous X-ray sources are detected
individually, sources below our detection limit are hidden in the
``diffuse'' emission. Indeed, the diffuse emission in the
cores of the \xs\ and \xq\ clusters shows a general correlation
with their stellar distributions (Fig.~\ref{f:rbp}). Thus,
relatively faint colliding wind binaries could significantly contribute
to the emission. But the bulk of the  diffuse X-ray
emission in outer regions of the clusters may have different
origins for several reasons. First, the emission extends much
further away from the cluster cores than the stellar light
distributions (Fig.~\ref{f:rbp}). Second, the spectrum of the
diffuse emission is harder than that of the discrete sources.
Third, the emission in the outer region of the \xs\ cluster mainly
exhibits the 6.4-keV line, inconsistent with the the colliding
wind interpretation.

\subsection{Cluster Winds}\label{ss:d_cw}

In addition to colliding winds in individual massive star
binaries, the collision among stellar winds collectively becomes
important in a compact cluster such as the \xs. The collision
results in the thermalization of the stellar winds and their
subsequent merging into a so-called cluster wind. Various 1-D
models and 3-D hydrodynamic simulations have been carried out on cluster winds
\citep{rag01,ste03,roc05}. 
Within the uncertainties of such model parameters as overall stellar wind velocities and mass loss rates, simulated cluster winds are shown to explain the luminosities of diffuse X-ray emission
from several star clusters \citep[e.g.,][]{ste03,roc05}; but little detailed comparison has yet been performed.

Fig.~\ref{f:rbp} compares the radial diffuse X-ray intensity profiles from the
3-D hydro-dynamical simulations, carried out specifically for the \xs\ and \xq\
clusters, approximately accounting for the discrete positions of
massive stars and their individual stellar wind properties
\citep{roc05}. For the \xq, the cluster
wind could account for $\sim 
1/4-1/3$ of the observed diffuse X-ray
emission. For the \xs, which is much more compact, the simulated
profile gives a reasonably good match to our measured distribution
of the diffuse X-ray intensity within $\sim 10$\as, but is too
steep to explain the emission at larger radii. The flattening of 
the observed intensity distribution in the radius range of
$\sim 10^{\prime\prime}$ to $\lesssim 15^{\prime\prime}$) may arise from
the reverse shock heating and confinement of the wind. At larger radii, the 
overall diffuse X-ray enhancement demonstrates a bow shock morphology and 
is prominent in the Fe K$\alpha$ 6.4-keV line emission
 (\S~\ref{sss:dif_a}), inconsistent with 
the expectation for the cluster wind interpretation (see below). Therefore, the
cluster wind may be important in the core, but not in the outer
region of the \xs\ cluster.

\begin{figure}[!thb]
  \centerline{
      \epsfig{figure=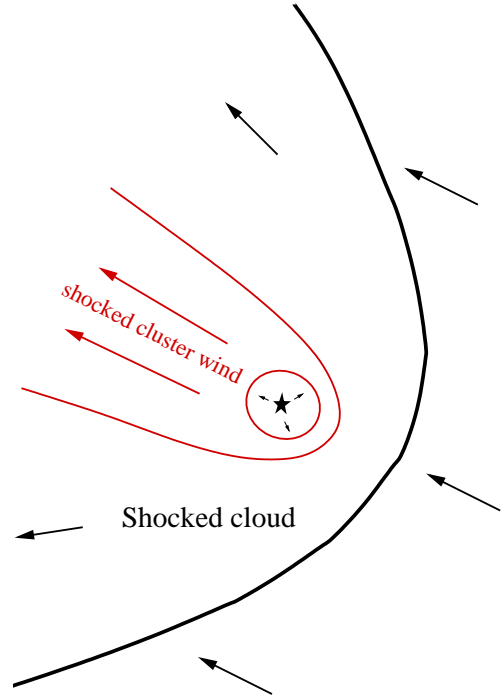,width=0.4\textwidth}
    }
\caption{\small  An illustration of the proposed cluster-cloud 
collision scenario for the \xs. The shocked cloud gas is partly traced by the CS 
and 6.4-keV lines (Fig.~\ref{f:cs_x}), whereas the shocked cluster
wind plasma near the cluster is by the 6.7-keV line (Fig.~\ref{f:im_line}). }
\label{f:ill}
\end{figure}

The complexity of the diffuse X-ray emission from the \xs\ cluster
probably reflects its interaction with the CS cloud. Both the morphology of
the diffuse X-ray emission, particularly the elongation of the 6.7-keV
line emission, and the extinction
deficit distribution indicate that the motion
of the cluster relative to the cloud is from East to West in the
sky.  Because of their supersonic relative motion, a bow-shock is expected to
form around the cluster. Fig.~\ref{f:ill} illustrates this  simple-minded 
scenario for the interaction, although the true situation is certainly
more complicated. 

 Following \citet{bur88}, we can estimate from the
ram-pressure balance the characteristic
 radius of the reverse shock in the cluster wind as
\begin{equation}
r_s = (0.7 {\rm~pc}) \dot{M}_{w,-4}^{1/2}
v_{w,3}^{1/2} v_{r,2}^{-1} n_{a,2}^{-1/2},
\label{e_r_s}
\end{equation}
where $\dot{M}_{w,-4}$ is the mass-loss rate of the cluster wind (in units of  $ 10^{-4} M_\odot$),
$v_{r,2}$ is the relative velocity between the cluster and the cloud ($10^2 {\rm~km~s^{-1}}$), and $n_{a,2}$ is the gas density
in the colliding cloud ($10^2 {\rm~cm^{-3}}$). Because the contact discontinuity has
a scale $l_c \sim 1.5 r_s$, we can estimate the volume of shocked wind
materials as $V \sim \frac{4\pi}{3}(1-1/1.5^3) l_c^3$.
Assuming that this volume corresponds to the 6.7-keV line plume,
which has a radius $l_c \sim 0.6$ pc (15\as) and
we can infer $n_e \sim 5 {\rm~cm^{-3}}$ from the integrated emission measure
of the central plume, $IEM \sim 16 {\rm~cm^{-6}~pc^3}$ (the
{\sl MEKAL} fit in Table~\ref{spec_d}).
The ram-pressure balance also gives the density of the shocked ambient
gas  $n_a \sim \frac{n_e}{4}(\frac{v_w}{v_{r}})^2$ $\sim
(1.3 \times 10^2  {\rm~cm^{-3}}) v_{w,3}^2 v_{r,2}^{-2}$. This, together with Eq.~\ref{e_r_s},
gives $\dot{M} \sim (1 \times 10^{-4} M_\odot) v_{w,3}$.

The above inferred $n_a$ and $\dot{M}$ values
depend  on $v_w$, which may be quite
uncertain. In particular, the near-IR spectroscopic estimate of
stellar winds may have significantly underestimated $v_w$ as
possible low-emissivity winds in the line profiles were not taken
into account \citep{cot96}, i.e., the wind terminal
velocity could be considerably higher than $1 \times 10^3
{\rm~km~s^{-1}}$. Nevertheless, the above inferred mass-loss
rate still appears substantially
smaller  (by a factor of up to $\sim 10$) than the current estimates based on radio continuum
estimates \citep[e.g.,][]{lan05}. Such estimates may be very uncertain 
(e.g., \citealt{roc05}), particularly for binaries with strong wind-wind interaction. 
The relatively small $n_a$ value is
consistent with the weak CS emission from the ambient
gas, probably representing the inter-clump medium of the colliding
cloud.

While the shocked cluster wind should be constantly flowing out
from the bow shock at a velocity comparable to the sound velocity
$c_s \sim (8 \times 10^2 {\rm~km~s^{-1}}) v_{w,3}$, we can also
estimate the ionization time scale as $\tau \sim n_e l_c/c_s \sim
(1 \times 10^{11} {\rm~cm^{-3}~s}) v_{w,3}^{-1}$, which
is much too large to explain the 6.4-keV line emission
with an {\sl NEI} plasma, but is consistent
with that inferred from the spectrum of the central plume
(\S~\ref{sss:dif_a}). Therefore, the observed
 size and shape of the 6.7-keV line plume (Fig.~\ref{f:im_line}),
at least qualitatively, match the
predictions of this simple bow shock interpretation, within the uncertainties
of the relevant parameters.

\subsection{Origin of the 6.4-keV line emission}\label{ss:d_line}

The above discussion indicates that the 6.4-keV line
emission associated with the \xs\ cluster is unlikely due to an {\sl NEI}
process. We thus consider the possible origin of the line emission
as the filling of iron K-shell vacancies
produced by either ionizing radiation with photon energies $> 7.1$ keV or
collision with low-energy cosmic-ray electrons
\citep[LECRe;][]{val00}. The fluorescent line emission and Thompson continuum
scattering seem to give a reasonable good explanation for those most
prominent 6.4-keV enhancements associated with well-known giant
molecular clouds such as Sgr B2 and Sgr C in the GC \citep{koy96a,
cra02, rev04}. This explanation requires the presence of
a luminous  X-ray source
with a spectrum consistent with the observed power law
continuum with a photon index of $\Gamma \approx 1.8$. Because
such a source is currently not present in the GC, the observed
emission is proposed to be the reflection of past Sgr A*,
with an X-ray luminosity of $\gtrsim 10^{39} {\rm~ergs~s^{-1}}$,
about a few hundreds years ago.

However, the fluorescence interpretation has difficulties in
accounting for the 6.4-keV line emission regions closer to Sgr A*.
A comparison of the CS emission and the diffuse 6.4-keV line
intensity does not show a peak-to-peak correlation, which should
be expected because the gas traced by the CS emission is expected
to be optically thin to the iron ionizing radiation \citep{wan03}.
As shown in \S~\ref{ss:cs_dist}, the detailed correlation
is also absent in the \xs\ CS cloud. This difficulty may be
avoided, if the CS emission does not trace well the
actual gas distribution (e.g., due to the destruction of the
molecule by the strong UV radiation from the \xs\ cluster). Even
in this case, however, the gas column density of the cloud cannot
be much greater than $\delta N_H \sim 10^{22} {\rm~cm^{-2}}$,
constrained by both the X-ray absorption and the near-IR
extinction distribution (\S~\ref{ss:d_envi}). Following
\citet{sun98}, we estimate the required X-ray
luminosity of Sgr A* to produce the detected 6.4-keV line
intensity of the \xs\ (Fig.~\ref{f:cs_x}) as
\begin{equation}
L_X = (4 \times 10^{39}{\rm~ergs~s^{-1}})
(d/27 {\rm~pc})^2 (\delta N_H/10^{22} {\rm~cm^{-2}})^{-1},
\end{equation}
where we have assumed the iron abundance to be 2 $\times$ solar and
have scaled the distance ($d$) between the cloud and Sgr A* to be
their projected separation in the sky, corresponding a light
travel time of only about 90 years. Of course, the actual distance
is likely to be greater, and the required $L_X$ should then be
higher. This common interpretation of the 6.4-keV line
enhancement and those associated with Sgr B2 and Sgr C, though
difficult to rule out completely, would not explain the apparent
position coincidence between  the cloud and the cluster.

Alternatively, one may consider the \xs\ cluster as the origin of
the hard X-rays. But this possibility can be easily dismissed
because of the absence of the 6.7-keV line (which is strong in
both the point-like sources and in the cluster core) in the
6.4-keV line enhancement. Furthermore, the observed X-ray luminosity of
the cluster is more than a factor of $10^2$ short of what is
required for the fluorescence interpretation.

A more plausible scenario for the \xs\ 6.4-keV line enhancement is
the LECRe-induced Fe K-shell vacancy filling \citep{val00}.
In this scenario, the continuum is due to the bremsstrahlung
radiation of the LECRe. The expected power-law photon index
of the continuum is 1.3-1.4 over
the range of 1-10 keV, consistent with our measured value of the
SE extension (Table~\ref{spec_d}). The
LECRe may be produced in strong shocks that are present within
the \xs\ cluster and in both the forward bow shock and the
reverse-shock in the cluster wind (see the discussion above). For
example, \citet{byk00} have shown that a shock of velocity
$\gtrsim 10^2 {\rm~km~s^{-1}}$ into a molecular cloud, accompanied
by magneto-hydrodynamic turbulence, can provide a spatially
inhomogeneous distribution of nonthermal LECRe. \citet{yus03} have
further presented observational evidence for nonthermal diffuse
radio emission from the \xs\ cluster and have suggested that
colliding wind shocks may generate the responsible relativistic
particles. The diffuse X-ray enhancement has a bow-shock
morphology and is presumably linked to the site of particle
acceleration. But, because of  particle
diffusion and gas flow, one  does not expect a
peak-to-peak correlation of the X-ray emission with the CS
emission from the colliding cloud. Following
\citet{yus02a}, we estimate the LECRe energy density required to produce the
observed 6.4-keV line intensity. If the {\sl shocked} gas
density is $\sim 10^{3} {\rm~cm^{-3}}$, the required energy
density is then $\sim 6 \times 10^3 {\rm~eV~cm^{-3}}$,
substantially greater than the value $0.2 {\rm~eV~cm^{-3}}$ from
averaging over the Galactic ridge \citep{val00}.
But the implied pressure inside the bow shock can still be
balanced by the high ram-pressure ($\sim 2 \times 10^{-8}
v_{r,2}^2 n_{a,2} {\rm~dyn~cm^{-2}}$) of the collision between the
cluster wind and the CS cloud. In short, the bow shock provides a
plausible interpretation of the distinct spatial and spectral
properties of the diffuse X-ray emission around the cluster and
its physical relationship to the CS cloud.

Finally, we consider the possibility that the 6.4-keV line enhancement represents
the reprocessed X-rays from numerous discrete and faint sources embedded
around the \xs\ cluster. A natural candidate for such sources might be low-mass
pre-main sequence young stellar objects (YSOs).
But they are in general not known to emit strong 6.4-keV line emission.
In the Orion nebula, for example, the line emission is detected from only a few
YSOs and with EWs less than 300 keV.
Therefore, YSOs are probably not a significant contributor to the 6.4-keV line
enhancement.

\subsection{YSO population and stellar IMF}\label{ss:d_yso}

The overall luminosity of the diffuse X-ray emission provides
a fundamental limit to the population of YSOs and hence the IMF
of the \xs\ and \xq\ clusters. YSOs in the mass range of
$(0.3 - 3) M_\odot$ typically have large
$L_X/L_{bol}$ ratios and hard X-ray spectra. Most importantly,
they can be numerous, as shown in the {\sl Chandra} Orion Ultra-deep Project
\citep{fei05}. Though with a {\sl mean} 2-8 keV luminosity
of only $\sim 1.2 \times 10^{30} {\rm~ergs~s^{-1}}$ per star, YSOs
collectively account for about 75\% of the luminosity of the 
Orion nebula,  the IMF of which is
consistent with the standard Miller \& Scalo (1979; MS hereafter), based on
the work by \citet{hil97}. If the clusters in the GC have a similar
IMF, YSOs should then be equally important.

At the GC, typical YSOs cannot be detected individually, but can be
constrained collectively in {\sl Chandra} observations.
Based on the diffuse X-ray intensity observed around Sgr A*,
\citet{nay05} find that the population of YSOs in the GC
cluster is extremely small. They conclude that
it cannot be a remnant of a massive star cluster, originated
at several tens of parsecs away from Sgr A* and then
dynamically spiralled in, and is thus most likely formed
{\sl in situ} in a self-gravitating circum-nuclear disk and
with a top heavy IMF. While star formation around a super-massive
black hole represents an extreme, it is clearly important
to examine the IMF of the \xs\ and \xq\ clusters
in the general environment of the GC.

\begin{figure}[htb]
  \centerline{
      \epsfig{figure=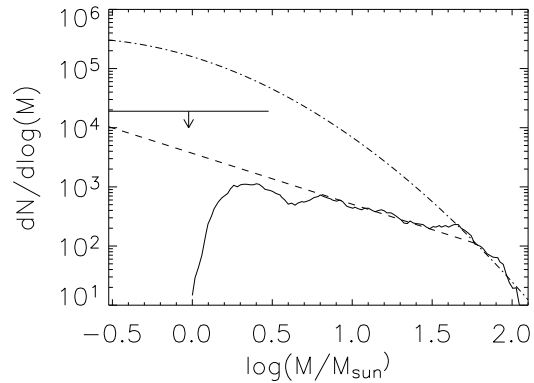,width=0.5\textwidth}
    }
\caption{\small Present-day MF as obtained by \citet{sto05} in
the $r < 0.4$~pc core of the \xs\ cluster, compared with 
the power law ($\propto M^\Gamma$, where $\Gamma=-0.86$; 
dashed curve) fitted in the 6-60 M$_\odot$
range \citep{sto05}) and the MS half-Gaussian (dot-dashed line), normalized to the number of stars in the  
mass range of $> 60$ M$_\odot$. The X-ray-inferred upper limit to the number of YSOs (0.3-3 M$_\odot$) 
in the entire cluster is marked as
the bar with the arrow. The MF at M $\lesssim$ 6 M$_\odot$ may be significantly contaminated by field
stars \citep{sto05}.}
\label{f:imf}
\end{figure}

We find a similar deficiency of YSOs in \xs\ and \xq\ clusters,
which places important constraints on their IMF. Existing near-IR
studies have provided estimates on the present-day MF of stars with
masses greater than a few solar masses in the core
of the \xs\ cluster; MF measurements in outer regions are difficult
because of severe confusion with field stars. 
Fig.~\ref{f:imf} shows the MF within $r < 0.4$ pc of the \xs\ cluster \citep{sto05},
compared with various predictions of YSOs.
The standard MS IMF, normalized with the number of stars 
in the mass range of $M > 60$ M$_\odot$, 
predicts at least 2$\times10^{5}$ YSOs. Using our
measurements of the diffuse X-ray emission, we can directly
get an upper limit to the population of YSOs over the entire cluster ($r < 2.5$ pc). We
assume that the mean X-ray luminosity of the YSOs in the \xs\
cluster to be the same as that in the Orion nebula, because of
their similar ages. As is shown above, much of the diffuse
X-ray emission, though difficult to quantify, likely has other
origins (e.g., cluster winds) to account for the prominent iron
K$\alpha$ lines. Therefore, the use of the total 2-8 keV diffuse
X-ray luminosity of $2 \times 10^{34} {\rm~ergs~s^{-1}}$ 
(Table~\ref{spec_d}) gives the upper limit as $2 \times
10^{4}$ YSOs, which is a factor of 10 smaller than the above
prediction from the MS IMF. The actual discrepancy should be substantially larger.
We have neglected the mass loss in the stellar evolution, which is important
for the massive stars. Considering the mass loss, the number of stars in the 
above initial mass range, hence the normalization of the IMF, would be greater.
The number of cluster stars in the region of $r= 0.4-2.5$ pc is also not included,
though difficult to fully quantify; for example, 
there are 77 stars with $M>40 M_{\odot}$ in the radius $r< $ 0.6 pc 
\citep{fig99b}, compared to about 48 in the same mass range of the 
MF obtained by \citet{sto05} for $r< $ 0.4 pc. 

An extrapolation of a power law fit in the mass range of 6-60 M$_\odot$, 
as obtained by \citet{sto05}, is
consistent with the X-ray-inferred number of YSOs (Fig.~\ref{f:imf}).
But, \citet{sto05} shows that the MF steepens with radius. This
steepening is expected as a result of the dynamic mass segregation
of stars in the cluster core \citep{kim02,por02}. Furthermore, mergers among stars may also be important in
flattening the MF toward the cluster core. Therefore, 
the MF of the entire cluster, including the region in $r = 0.4 - 2.5$ pc, is 
likely to be steeper. If this is the case, a 
turnover in the MF (e.g., at $\sim 6$ M$_{\odot}$, as
indicated in the study of \citet{sto05}, may indeed be required to
explain the X-ray-inferred upper limit on the overall YSO population in the 
\xs\ cluster.

Similarly, we can constrain the YSO population in the \xq\
cluster. There are 30 stars with masses larger than 20 $M_{\odot}$
within $r = 25$\as\ (1 pc) of the \xq\ cluster \citep{fig99a}. Assuming
the MS IMF would predict a total number of YSOs to be at least
$2 \times10^{4}$. These YSOs would have a 2-8 keV luminosity of
$\sim 1 \times10^{34} {\rm~ergs~s^{-1}}$, accounting for the
weak dependence of the mean X-ray luminosity on the cluster age (a
factor of 1.6; \citealt{pre05}). This predicted value is a
factor of 5 greater than our measured total diffuse X-ray luminosity of $2 \times10^{33}
{\rm~ergs~s^{-1}}$ within $\sim 1^\prime$ of the \xs2\ cluster (\S~\ref{sss:dif_a}).

\section{Summary}

We have presented a deep \chandra\ ACIS-I observation of the
\xs\ and \xq\ clusters. This observation, complemented by a high-resolution
OVRO mapping of a CS cloud accompanied with the \xs\ cluster, allows for an in-depth
study of the high-energy phenomena and processes in these two clusters and their
interplay with the GC environment. The main results of our study are as follows:
\begin{itemize}

\item Point-like X-ray sources in the \chandra\ observation are detected 
down to a detection limit of $ \sim 5 \times 10^{31}
{\rm~ergs~s^{-1}}$ in the 1-9 keV band.
This list should contain all significant massive colliding wind binaries in the
ACIS-I field. The superb \chandra\ positioning of
these sources allow for identifications of multi-wavelength counterparts.
We consider those sources best-detected in the 1-9 keV band to be likely
located in the GC, whereas those sources preferentially detected in the 1-4 keV or
4-9 keV band are good candidates for foreground stars or extragalactic AGNs.
In particular, we estimate that the AGN contribution is on the order of only
a few \% of all our detected X-ray sources.

\item The X-ray source number-flux relation
of the \xs\ and \xq\ cluster region can be characterized
by a power law $N(<S) = N_0(S/S_0)^{-\alpha}$, where
$N_0 = 0.14
{\rm~sources~arcmin^{-2}}$, $\alpha =1.26_{-0.13}^{+0.14}$, and
$S_0 = 3 \times 10^{-6} {\rm~ph~cm^{-2}~s^{-1}}$, equivalent to $6
\times 10^{-14}$$ {\rm~ergs~cm^{-2}~s^{-1}}$, in the 0.5-8 keV band.
This relation is significantly flatter than those determined in
 other regions of the GC \citep{mun06}, apparently due to the
presence of a massive star-related  population of relatively
bright X-ray sources.

\item The three bright point-like X-ray sources in the core of the
Arches cluster show remarkably similar spectra, which can all be
characterized by an optically-thin thermal plasma with a temperature
of $\sim 1.8-2.5$ keV, a metal abundance of $\sim
1.8^{+0.8}_{-0.2}$ solar, and a foreground absorption of
7.7$^{+0.8}_{-0.8}\times10^{22}$ ${\rm~cm^{-2}}$. The 0.3-8 keV
luminosities of the sources are in the range of $(5-11)\times 10^{33}$
${\rm~ergs~s^{-1}}$. The sources have near-IR counterparts as late-type WN
stars, which tend to have massive and fast stellar winds.
Therefore, the sources likely represent the extreme examples of
colliding stellar wind binaries.
The measured super-solar metal (iron)
abundance is consistent with other X-ray measurements of thermal
hot plasma in the GC environment.

\item The X-ray sources in the core of the older and looser \xq\
cluster are substantially dimmer and show more diverse spectral
characteristics. QX1, with a soft spectrum, is clearly a
foreground star, whereas QX4, with a very hard X-ray spectrum, but
without a near-IR counterpart, could be either a background AGN or
a strongly obscured stellar object. The remaining two (QX2 and
QX3) are probably colliding wind binary systems, albeit less
energetic than those in the \xs\ cluster. The Pistol star, despite
of its enormous bolometric luminosity, is not detected with a
3$\sigma$ upper limit to the 0.3-8 keV luminosity as 3$\times10^{33}$
${\rm~ergs~s^{-1}}$, consistent with the norminal relation
$L_{x}/L_{bol} \sim 10^{-7}$.

\item Diffuse X-ray emission from
both \xs\ and \xq\ clusters are unambiguously detected. The emission is substantially more
extended than the stellar distributions. The emission in the core
region of \xs\ ($r \lesssim 0.6$ pc) has a 2-8 keV
luminosity of $4 \times 10^{33} {\rm~ergs~s^{-1}}$, a steep radial
intensity decline, and a hard spectrum with a strong highly-ionized Fe K$\alpha$ line.
These properties are consistent with the cluster wind interpretation.

\item The diffuse X-ray emission outside the \xs\ core, however,
has a relatively flat and
non-axis-symmetric spatial distribution. The spectrum of the emission shows a
distinct 6.4-keV line with an EW of $\sim 1.4$ keV. This line cannot be
explained by the fluorescence of the \xs\ cluster X-ray emission and is probably
due to the collision of low-energy cosmic-ray electrons with neutral
or weakly ionized irons in a bow shock, which results from the
supersonic motion of the cluster relative to the CS cloud.

\item  The diffuse X-ray emission from the \xq\ cluster ($L_X \sim 
2 \times 10^{33} {\rm~ergs~s^{-1}}$) is about a factor of 10 lower than
from the \xs\ cluster and can be naturally explained by the cluster wind and
 a limited number of low-mass pre-main sequence YSOs.

\item There appears to be a general deficiency of YSOs in the two
clusters, relative to the prediction from 
the standard Miller \& Scalo IMF. Compared with the X-ray emission from young stars in the
Orion nebula, our observed total diffuse X-ray luminosities from
the Arches and Quintuplet clusters suggest that they contain no
more than $2 \times 10^4$ and  $3 \times 10^3$ YSOs. These numbers are 
a factor of 10 and 5 smaller than what would be expected from the
IMF and the massive star popluations observed in the cores of the two clusters. 
One possibility is that the IMF indeed flattens at intermediate masses,
as indicated in a near-IR study of the inner regions of the \xs\ 
cluster \citep{sto05}. 

\item The CS molecular cloud appears to be colliding with the \xs\ cluster at
a relative velocity of $\gtrsim 120 {\rm~km~s^{-1}}$, explaining the bow shock
morphology of the associated diffuse X-ray emission and its weak correlation
with the CS (J = 2--1) line intensity as well as the near-IR extinction
distribution and possibly the abnormal kinematics of the  
dusty gas in the field. Such collisions may be responsible for 
removing the natal clouds of the clusters at their early ages. High-velocity
cloud-cloud collisions might also initiate the formation of massive 
stellar clusters, such 
as the \xs\ and \xs2, with top-heavy IMFs. These are the effects unique 
in the Galactic nuclear environment.

\end{itemize}

\acknowledgements We thank D. Figer for sending us near-IR data for
comparison and for useful discussion on the Arches cluster,
A. Stolte for her comments on the MF determination, and the 
anonymous referee for useful comments and suggestions. This work
is supported by NASA through the grant SAO/CXC GO4-5010X.

\vfil
\end{document}